# Anisotropy in Double Negative Permittivity and Permeability Materials or Left-Handed Materials — The Effect on Guided Wave Propagating Electromagnetic Fields

Clifford M. Krowne

Microwave Technology Branch, Electronics Science & Technology Division,
Naval Research Laboratory, Washington, DC 20375

ABSTRACT

Effect of anisotropy in the physical tensor description of the negative index of refraction material acting as a substrate is found on the electromagnetic field distributions. This is done for the case of a microstrip structure whose configuration is commonly used in microwave and millimeter wave integrated circuits. These ab initio studies have been done self-consistently with a computer code using a full-wave integral equation numerical method based upon a generalized Green's function utilizing appropriate boundary conditions. Field distributions are provided over two decades of frequency in the cross-section of the uniform guided wave structure, from 0.2 GHz to 20 GHz. It is found that modifying the tensor can allow control because the wave changes volumetrically, or switches from volumetric to surface, in its distribution of fields.

# I. INTRODUCTION

Although considerable physics has been learned about the dispersion behavior and electromagnetic fields for propagation down guided wave microstrip structures loaded with DNM (double negative materials with simultaneous negative permittivity and permeability) which are also referred to LHM (left-handed-materials for the left-handed orientation of the electric, magnetic, and propagation phase constant) [1], [2] nothing is known about what happens to the field distributions as anisotropy is introduced. This is an extremely interesting area since maintenance of isotropy has been recognized to be essential for 3D imaging possibilities ([3], [4]), and such isotropy should also be necessary for arbitrary field contouring or arrangement of fields in small electronic devices which are created especially to employ the unique properties of DNM. It should also be noted here that these materials can be referred to as NIRM (negative index of refraction materials) or NIM (negative index materials). Furthermore, because the essential property of these materials is opposite orientations of the phase velocity and the Poynting vector (giving the power flow direction), they may be referred to as alternatively as negative phase velocity materials (NPV materials or simply NPVM) [5].

Because we are not going to further examine 3D imaging issues, but 3D behavior of fields in electronic devices, we will restrict the use of notation when referring to these materials as DNM or NPVM. Although maintaining isotropy for many integrated circuit applications may be desirable as mentioned above, it may also be desirable to introduce anisotropy ([6], [7]), or at least find it acceptable to have some anisotropy in certain device configurations, where because of the particular applications, the effective dimensions of the device is reduced from 3D to 2D or even 1D. This is not some idle speculation, as quasi-lumped element realizations using distributed and/or lumped sections have been used to make components employing backward wave behavior, one of the hallmarks of DNM or NPVM [8]-[11]. Thus, whether or not we want to examine what the effect is of some deviation from isotropy, or wish to intentionally introduce anisotropy, it would be very instructive to undertake such an investigation.

In pursuing this quest, the backdrop of already studied field distributions for isotropic DNM in the C-band, X-band, Ka and V-bands, and W-band broadcast frequency regions, gives us some basis upon which to begin these studies. In the C-band region, electromagnetic field line plots of $\mathbf{E}_t$ and $\mathbf{H}_t$ for the transverse fields in the cross-section perpendicular to the propagation direction were provided at 5 GHz [2]. Also given were simultaneous magnitude and arrow vector distributions plots for E/ $\mathbf{E}_t$ and H/ $\mathbf{H}_t$ at the same frequency. Finally, a Poynting vector distribution plot $P_z$ was provided at 5 GHz. In the X-band region, electromagnetic field line plots of $\mathbf{E}_t$ and $\mathbf{H}_t$ were provided at 10 GHz [12]. Also given were simultaneous magnitude and arrow vector distributions plots for E/ $\mathbf{E}_t$ and H/ $\mathbf{H}_t$ at the same frequency. Finally, a Poynting vector distribution plot $P_z$ was provided at 10 GHz. In the overlap region between the Ka and V-bands, electromagnetic field plots of simultaneous magnitude and arrow vector distributions plots for E/ $\mathbf{E}_t$ and H/ $\mathbf{H}_t$ were given at 40 GHz [12], [13]. Simultaneous line and magnitude distribution plots for E/ $\mathbf{E}_t$ and H/ $\mathbf{H}_t$ were also provided at 40GHz [12]. Lastly, a Poynting vector distribution plot $P_z$ was provided at the same frequency [12].

The following two sections will cover the governing equations (Section II), acquisition of the anisotropic Green's function (Section II), use of basis current functions (Section III), and determination of the propagation constant and electromagnetic fields (Section III) for the guided wave microstrip structure with anisotropic DNM. With a complete development in hand, numerical calculations are performed in Section IV for a microstrip structure at three frequencies offset from each other by decade steps. Each study is begun by first examining an isotropic tensor to provide a reference standard for looking at the effects of introducing anisotropy through a permittivity tensor. Once the electromagnetic distributions have been obtained for the isotropic case, distributions for anisotropy are calculated. This process of first finding the isotropic result, then proceeding on to the anisotropic situation, is done at each frequency.

## II. ANISOTROPIC GREEN'S FUNCTION BASED UPON LHM or DNM PROPERTIES

Maxwell's time varying equations describe the electromagnetic field behavior in a medium if they are combined with constitutive relationships embedding the physical properties of the medium in them. Maxwell's two curl equations are

$$\nabla \times \mathbf{E}(t,\mathbf{x}) = -\frac{\partial \mathbf{B}(t,\mathbf{x})}{\partial t} \quad ; \quad \nabla \times \mathbf{H}(t,\mathbf{x}) = \frac{\partial \mathbf{D}(t,\mathbf{x})}{\partial t} + \mathbf{J}(t,\mathbf{x}) \tag{1}$$

Constitutive relationships are

$$\mathbf{D}(t,\mathbf{x}) = \bar{\bar{\varepsilon}}\mathbf{E}(t,\mathbf{x}) + \bar{\bar{\rho}}\mathbf{H}(t,\mathbf{x}) \quad ; \quad \mathbf{B}(t,\mathbf{x}) = \bar{\bar{\rho}}'\mathbf{E}(t,\mathbf{x}) + \bar{\bar{\mu}}\mathbf{H}(t,\mathbf{x}) \tag{2}$$

Here $\mathbf{x} = (x_1, x_2, x_3) = (x, y, z)$. Most general NPV medium can have all constitutive tensors in (2) nonzero, including the magnetoelectric or optical activity tensors $\bar{\bar{\rho}}$ and $\bar{\bar{\rho}}'$, as they are sometimes called. The formulation is therefore kept general in order to retain the most flexibility for studying materials with widely varying physical properties. Because many problems are often transparent in the frequency domain, and because nonharmonic problems can be resolved into a superposition of time-harmonic components, we elect to study here time harmonic electromagnetic wave propagation through the solid state LHM-RHM structure (RHM = right-handed-medium, or ordinary medium). Taking the time harmonic variation to be of a form $e^{i\omega t}$, Maxwell's equation become

$$\nabla \times \mathbf{E}(\mathbf{x}) = -i\omega \mathbf{B}(\mathbf{x}) \quad ; \quad \nabla \times \mathbf{H}(\mathbf{x}) = i\omega \mathbf{D}(\mathbf{x}) + \mathbf{J}(\mathbf{x}) \tag{3}$$

with the constitutive relationships dropping the explicit t-dependence

$$\mathbf{D}(\mathbf{x}) = \bar{\bar{\varepsilon}}(\omega)\mathbf{E}(\mathbf{x}) + \bar{\bar{\rho}}(\omega)\mathbf{H}(\mathbf{x}) \quad ; \quad \mathbf{B}(\mathbf{x}) = \bar{\bar{\rho}}'(\omega)\mathbf{E}(\mathbf{x}) + \bar{\bar{\mu}}(\omega)\mathbf{H}(\mathbf{x}) \tag{4}$$

Dependence of the constitutive parameters on radian frequency is a well recognized fact and that is why explicit variation on $\omega$ is shown. However, for the study to be conducted here at specific frequencies, we do not need to call out this dependence explicitly. We will be setting for, example $\bar{\bar{\varepsilon}}(\omega) = v$ for $\omega = \omega_v$. Therefore, we set

$$\mathbf{D}(\mathbf{x}) = \bar{\bar{\varepsilon}}\mathbf{E}(\mathbf{x}) + \bar{\bar{\rho}}\mathbf{H}(\mathbf{x}) \quad ; \quad \mathbf{B}(\mathbf{x}) = \bar{\bar{\rho}}'\mathbf{E}(\mathbf{x}) + \bar{\bar{\mu}}\mathbf{H}(\mathbf{x}) \tag{5}$$

and understand it means (2).

Curl equations (1) may be combined into a single sourceless governing equation [14],

$$L_T(\mathbf{x})\mathbf{V}_L(\mathbf{x}) = i\omega \mathbf{V}_R(\mathbf{x}) \tag{6}$$

where the matrix partial differential operator acting on the **E-H** column vector is

$$L_T(\mathbf{x}) = \begin{bmatrix} 0 & L_q(\mathbf{x}) \\ -L_q(\mathbf{x}) & 0 \end{bmatrix} \quad (7)$$

where the quadrant operator is

$$L_q(\mathbf{x}) = \begin{bmatrix} 0 & -\frac{\partial}{\partial z} & \frac{\partial}{\partial y} \\ \frac{\partial}{\partial z} & 0 & -\frac{\partial}{\partial x} \\ -\frac{\partial}{\partial y} & \frac{\partial}{\partial x} & 0 \end{bmatrix} \quad (8)$$

Current **J** effects are introduced later through a Green's function process [see (80)]. Vectors in (6) are

$$\mathbf{V}_L(\mathbf{x}) = \begin{bmatrix} E_x \\ E_y \\ E_z \\ H_x \\ H_y \\ H_z \end{bmatrix} = \begin{bmatrix} \mathbf{E} \\ \mathbf{H} \end{bmatrix} \quad ; \quad \mathbf{V}_R(\mathbf{x}) = \begin{bmatrix} D_x \\ D_y \\ D_z \\ B_x \\ B_y \\ B_z \end{bmatrix} = \begin{bmatrix} \mathbf{D} \\ \mathbf{B} \end{bmatrix} \quad (9)$$

Restricting ourselves to a guided wave structure with the wave traveling in a uniform cross-section in the z-direction, that is the wave has the form $e^{i\omega t - \gamma z}$, $\gamma = \gamma(\omega)$, simplifies (6)-(8) to

$$L_T(x,y)\mathbf{V}_L(x,y) = i\omega \mathbf{V}_R(x,y) \quad (10)$$

$$L_T(x,y) = \begin{bmatrix} 0 & L_q(x,y) \\ -L_q(x,y) & 0 \end{bmatrix} \quad (11)$$

$$L_q(x,y) = \begin{bmatrix} 0 & \gamma & \frac{\partial}{\partial y} \\ -\gamma & 0 & -\frac{\partial}{\partial x} \\ -\frac{\partial}{\partial y} & \frac{\partial}{\partial x} & 0 \end{bmatrix} \quad (12)$$

Finally, there are certain advantages for approaching the problem in the Fourier transform domain, not the least of which is that real space convolution integrals reduce to products, and so for the integral equation technique to be applied to a finite enclosed region in the x-direction, with layering in the y-direction, the finite Fourier transform pair

$$\mathbf{F}(k_x,y) = \int_{-b}^{b} \mathbf{F}(x,y)e^{-ik_x x}dx \quad ; \quad F_i(k_x,y) = \int_{-b}^{b} F_i(x,y)e^{-ik_x x}dx \qquad (13)$$

$$\mathbf{F}(x,y) = \frac{1}{2b}\sum_{k_x=-\infty}^{\infty}\mathbf{F}(k_x,y)e^{ik_x x} \quad ; \quad F_i(x,y) = \frac{1}{2b}\sum_{k_x=-\infty}^{\infty}F_i(k_x,y)e^{ik_x x} \qquad (14)$$

is applied to the fields, converting (10) — (12) into

$$L_T(k_x,y)\mathbf{V}_L(k_x,y) = i\omega \mathbf{V}_R(k_x,y) \qquad (15)$$

$$L_T(k_x,y) = \begin{bmatrix} 0 & L_q(k_x,y) \\ -L_q(k_x,y) & 0 \end{bmatrix} \qquad (16)$$

$$L_q(k_x,y) = \begin{bmatrix} 0 & \gamma & \dfrac{\partial}{\partial y} \\ -\gamma & 0 & -ik_x \\ -\dfrac{\partial}{\partial y} & ik_x & 0 \end{bmatrix} \qquad (17)$$

Constitutive relationships (2) can be combined into a super-tensor,

$$\mathbf{V}_R(t,\mathbf{x}) = \mathbf{M}\mathbf{V}_L(t,\mathbf{x}) \quad ; \quad \mathbf{M} = \begin{bmatrix} \bar{\bar{\varepsilon}} & \bar{\bar{\rho}} \\ \bar{\bar{\rho}}' & \bar{\bar{\mu}} \end{bmatrix} \qquad (18)$$

and using the harmonic transformation leading to (5),

$$\mathbf{V}_R(\mathbf{x}) = \mathbf{M}(\omega)\mathbf{V}_L(\mathbf{x}) \qquad (19)$$

Characterization of the wave by the complex propagation constant reduces (19) to

$$\mathbf{V}_R(x,y) = \mathbf{M}(\omega)\mathbf{V}_L(x,y) \qquad (20)$$

When the finite Fourier transform is applied to (20),

$$\mathbf{V}_R(k_x,y) = \mathbf{M}(\omega)\mathbf{V}_L(k_x,y) \qquad (21)$$

Inserting this formula describing the material physics into electromagnetic equation yields, after eliminating $\mathbf{V}_R$,

$$L_T(k_x,y)\mathbf{V}_L(k_x,y) = i\omega \mathbf{M}(\omega)\mathbf{V}_L(k_x,y) \qquad (22)$$

This matrix equation can in principle be solved for the normal mode eigenvectors and eigenvalues $\gamma = \gamma(k_x, \omega)$, realizing that the 1×6 column vectors and the 6×6 operator and material tensor square matrices use the full electromagnetic field component set. However, because we will restrict ourselves to canonical layered structures (layered in the y-direction), it is very convenient to extract the perpendicular field components from (22) using rows 2 and 5 which do not possess differential operator d/dy:

$$-\gamma V_4 - ik_x V_6 = i\omega \sum_{i=1}^{6} m_{2i} V_i \qquad (23)$$

$$\gamma V_1 + ik_x V_3 = i\omega \sum_{i=1}^{6} m_{5i} V_i \qquad (24)$$

Solution of (23) and (24) is

$$V_i = \sum_{j=1}^{6} a_{ij}(1-\delta_{2,j})(1-\delta_{5,j})V_j \quad ; \quad i = 2,5 \qquad (25)$$

where $\delta_{i,j}$ is the Kronecker delta function, or

$$E_y = a_{21}E_x + a_{23}E_z + a_{24}H_x + a_{26}H_z \qquad (26)$$

$$H_y = a_{51}E_x + a_{53}E_z + a_{54}H_x + a_{56}H_z \qquad (27)$$

Here $a_{ij}$ are given by

$$a_{ij} = \frac{a'_{ij}}{D_a} \qquad (28)$$

$$D_a = m_{22}m_{55} - m_{25}m_{52} \qquad (29)$$

$$a'_{21} = m_{25}\left(m_{51} + \frac{i\gamma}{\omega}\right) - m_{21}m_{55} \qquad (30)$$

$$a'_{23} = m_{25}\left(m_{53} - \frac{k_x}{\omega}\right) - m_{23}m_{55} \qquad (31)$$

$$a'_{24} = m_{25}m_{54} - m_{55}\left(m_{24} - \frac{i\gamma}{\omega}\right) \qquad (32)$$

$$a'_{26} = m_{25}m_{56} - m_{55}\left(m_{26} + \frac{k_x}{\omega}\right) \qquad (33)$$

$$a'_{51} = m_{52}m_{21} - m_{22}\left(m_{51} + \frac{i\gamma}{\omega}\right) \qquad (34)$$

$$a'_{53} = m_{52}m_{23} - m_{22}\left(m_{53} - \frac{k_x}{\omega}\right) \qquad (35)$$

$$a'_{54} = m_{52}\left(m_{24} - \frac{i\gamma}{\omega}\right) - m_{22}m_{54} \tag{36}$$

$$a'_{56} = m_{52}\left(m_{26} + \frac{k_x}{\omega}\right) - m_{22}m_{56} \tag{37}$$

Here **M** consists of the set $\{m_{ij}\}$ of elements. For the case when the optical activities are turned off, $\bar{\bar{\rho}} = 0$ and $\bar{\bar{\rho}}' = 0$, (29) — (37) become

$$D_a = \varepsilon_{22}\mu_{22} - \rho_{22}\rho'_{22} = \varepsilon_{22}\mu_{22} \tag{38}$$

$$a'_{21} = \rho_{22}\left(\rho'_{21} + \frac{i\gamma}{\omega}\right) - \varepsilon_{21}\mu_{22} = -\varepsilon_{21}\mu_{22} \tag{39}$$

$$a'_{23} = \rho_{22}\left(\rho'_{23} - \frac{k_x}{\omega}\right) - \varepsilon_{23}\mu_{22} = -\varepsilon_{23}\mu_{22} \tag{40}$$

$$a'_{24} = \rho_{22}\mu_{21} - \mu_{22}\left(\rho_{21} - \frac{i\gamma}{\omega}\right) = \mu_{22}\frac{i\gamma}{\omega} \tag{41}$$

$$a'_{26} = \rho_{22}\mu_{23} - \mu_{22}\left(\rho_{23} + \frac{k_x}{\omega}\right) = -\mu_{22}\frac{k_x}{\omega} \tag{42}$$

$$a'_{51} = \rho'_{22}\varepsilon_{21} - \varepsilon_{22}\left(\rho'_{21} + \frac{i\gamma}{\omega}\right) = -\varepsilon_{22}\frac{i\gamma}{\omega} \tag{43}$$

$$a'_{53} = \rho'_{22}\varepsilon_{23} - \varepsilon_{22}\left(\rho'_{23} - \frac{k_x}{\omega}\right) = \varepsilon_{22}\frac{k_x}{\omega} \tag{44}$$

$$a'_{54} = \rho'_{22}\left(\rho_{21} - \frac{i\gamma}{\omega}\right) - \varepsilon_{22}\mu_{21} = -\varepsilon_{22}\mu_{21} \tag{45}$$

$$a'_{56} = \rho'_{22}\left(\rho_{23} + \frac{k_x}{\omega}\right) - \varepsilon_{22}\mu_{23} = -\varepsilon_{22}\mu_{23} \tag{46}$$

For biaxial electric and magnetic crystalline properties in a principal axis system, only the diagonal elements of the sub tensors of M survive, making (39) — (46) become

$$a'_{21} = a'_{23} = 0 \,;\, a'_{24} = \mu_{22}\frac{i\gamma}{\omega} \,;\, a'_{26} = -\mu_{22}\frac{k_x}{\omega} \tag{47}$$

$$a'_{51} = -\varepsilon_{22}\frac{i\gamma}{\omega} \,;\, a'_{53} = \varepsilon_{22}\frac{k_x}{\omega} \,;\, a'_{54} = a'_{56} = 0 \tag{48}$$

Governing equation of the problem can be acquired by realizing that rows 1, 3, 4, and 6 of (22) contain first order linear differential equations

$$W_5 + \frac{dV_6}{dy} = i\omega\sum_{i=1}^{6} m_{1i}V_i \tag{49}$$

$$-\frac{dV_4}{dy} + ik_xV_5 = i\omega\sum_{i=1}^{6} m_{3i}V_i \tag{50}$$

$$-W_2 - \frac{dV_3}{dy} = i\omega\sum_{i=1}^{6} m_{4i}V_i \tag{51}$$

$$\frac{dV_1}{dy} - ik_xV_2 = i\omega\sum_{i=1}^{6} m_{6i}V_i \tag{52}$$

With the help of (25) to remove $V_2$ and $V_5$ from (49)-(52),

$$\frac{d\aleph}{dy} = i\omega \mathbf{R} \aleph \tag{53}$$

$$\aleph = \begin{bmatrix} V_1 \\ V_3 \\ V_4 \\ V_6 \end{bmatrix} = \begin{bmatrix} E_x \\ E_z \\ H_x \\ H_z \end{bmatrix} \tag{54}$$

$$r_{1j} = m_{6\theta} + a_{5\theta} m_{65} + a_{2\theta}\left(m_{62} + \frac{k_x}{\omega}\right) \tag{55}$$

$$r_{2j} = -\left\{m_{4\theta} + a_{5\theta} m_{45} + a_{2\theta}\left(m_{42} - \frac{i\gamma}{\omega}\right)\right\} \tag{56}$$

$$r_{3j} = -\left\{m_{3\theta} + a_{2\theta} m_{32} + a_{5\theta}\left(m_{35} - \frac{k_x}{\omega}\right)\right\} \tag{57}$$

$$r_{4j} = m_{1\theta} + a_{2\theta} m_{12} + a_{5\theta}\left(m_{15} + \frac{i\gamma}{\omega}\right) \tag{58}$$

Here the $\theta(j)$ index on $a_{ij}$ is defined by

$$\theta(j) = \begin{cases} \dfrac{3}{2} j & ; \quad j = 2, 4 \\ \dfrac{3j-1}{2} & ; \quad j = 1, 3 \end{cases} \tag{59}$$

Here $\mathbf{R}$ consists of the set $\{r_{ij}\}$ of elements. For the case when the optical activities are turned off, $\overline{\overline{\rho}} = 0$ and $\overline{\overline{\rho}}' = 0$,

$$r_{11} = \rho'_{31} + a_{51}\mu_{32} + a_{21}\left(\rho'_{32} + \frac{k_x}{\omega}\right) = -\frac{i\gamma}{\omega}\frac{\mu_{32}}{\mu_{22}} - \frac{\varepsilon_{21}}{\varepsilon_{22}}\frac{k_x}{\omega} \tag{60}$$

$$r_{12} = \rho'_{33} + a_{53}\mu_{32} + a_{23}\left(\rho'_{32} + \frac{k_x}{\omega}\right) = \frac{k_x}{\omega}\frac{\mu_{32}}{\mu_{22}} - \frac{\varepsilon_{23}}{\varepsilon_{22}}\frac{k_x}{\omega} = \frac{k_x}{\omega}\left(\frac{\mu_{32}}{\mu_{22}} - \frac{\varepsilon_{23}}{\varepsilon_{22}}\right) \tag{61}$$

$$r_{13} = \mu_{31} + a_{54}\mu_{32} + a_{24}\left(\rho'_{32} + \frac{k_x}{\omega}\right) = \mu_{31} - \frac{\mu_{21}}{\mu_{22}}\mu_{32} + \frac{1}{\varepsilon_{22}}\frac{i\gamma}{\omega}\frac{k_x}{\omega} \tag{62}$$

$$r_{14} = \mu_{33} + a_{56}\mu_{32} + a_{26}\left(\rho'_{32} + \frac{k_x}{\omega}\right) = \mu_{33} - \frac{\mu_{23}}{\mu_{22}}\mu_{32} - \frac{1}{\varepsilon_{22}}\frac{k_x}{\omega}\frac{k_x}{\omega} \tag{63}$$

$$r_{21} = -\left\{\rho'_{11} + a_{51}\mu_{12} + a_{21}\left(\rho'_{12} - \frac{i\gamma}{\omega}\right)\right\} = \frac{i\gamma}{\omega}\frac{\mu_{12}}{\mu_{22}} - \frac{\varepsilon_{21}}{\varepsilon_{22}}\frac{i\gamma}{\omega} = \frac{i\gamma}{\omega}\left(\frac{\mu_{12}}{\mu_{22}} - \frac{\varepsilon_{21}}{\varepsilon_{22}}\right) \tag{64}$$

$$r_{22} = -\left\{\rho'_{13} + a_{53}\mu_{12} + a_{23}\left(\rho'_{12} - \frac{i\gamma}{\omega}\right)\right\} = -\frac{k_x}{\omega}\frac{\mu_{12}}{\mu_{22}} - \frac{\varepsilon_{23}}{\varepsilon_{22}}\frac{i\gamma}{\omega} \tag{65}$$

$$r_{23} = -\left\{\mu_{11} + a_{54}\mu_{12} + a_{24}\left(\rho'_{12} - \frac{i\gamma}{\omega}\right)\right\} = -\mu_{11} + \frac{\mu_{21}}{\mu_{22}}\mu_{12} + \frac{1}{\varepsilon_{22}}\frac{i\gamma}{\omega}\frac{i\gamma}{\omega} \tag{66}$$

$$r_{24} = -\left\{\mu_{13} + a_{56}\mu_{12} + a_{26}\left(\rho'_{12} - \frac{i\gamma}{\omega}\right)\right\} = -\mu_{13} + \frac{\mu_{23}}{\mu_{22}}\mu_{12} - \frac{1}{\varepsilon_{22}}\frac{k_x}{\omega}\frac{i\gamma}{\omega} \tag{67}$$

$$r_{31} = -\left\{\varepsilon_{31} + a_{21}\varepsilon_{32} + a_{51}\left(\rho_{32} - \frac{k_x}{\omega}\right)\right\} = -\varepsilon_{31} + \frac{\varepsilon_{21}}{\varepsilon_{22}}\varepsilon_{32} - \frac{1}{\mu_{22}}\frac{i\gamma}{\omega}\frac{k_x}{\omega} \tag{68}$$

$$r_{32} = -\left\{\varepsilon_{33} + a_{23}\varepsilon_{32} + a_{53}\left(\rho_{32} - \frac{k_x}{\omega}\right)\right\} = -\varepsilon_{33} + \frac{\varepsilon_{23}}{\varepsilon_{22}}\varepsilon_{32} + \frac{1}{\mu_{22}}\frac{k_x}{\omega}\frac{k_x}{\omega} \tag{69}$$

$$r_{33} = -\left\{\rho_{31} + a_{24}\varepsilon_{32} + a_{54}\left(\rho_{32} - \frac{k_x}{\omega}\right)\right\} = -\frac{i\gamma}{\omega}\frac{\varepsilon_{32}}{\varepsilon_{22}} - \frac{\mu_{21}}{\mu_{22}}\frac{k_x}{\omega} \tag{70}$$

$$r_{34} = -\left\{\rho_{33} + a_{26}\varepsilon_{32} + a_{56}\left(\rho_{32} - \frac{k_x}{\omega}\right)\right\} = \frac{k_x}{\omega}\frac{\varepsilon_{32}}{\varepsilon_{22}} - \frac{\mu_{23}}{\mu_{22}}\frac{k_x}{\omega} = \frac{k_x}{\omega}\left(\frac{\varepsilon_{32}}{\varepsilon_{22}} - \frac{\mu_{23}}{\mu_{22}}\right) \tag{71}$$

$$r_{41} = \varepsilon_{11} + a_{21}\varepsilon_{12} + a_{51}\left(\rho_{12} + \frac{i\gamma}{\omega}\right) = \varepsilon_{11} - \frac{\varepsilon_{21}}{\varepsilon_{22}}\varepsilon_{12} - \frac{1}{\mu_{22}}\frac{i\gamma}{\omega}\frac{i\gamma}{\omega} \tag{72}$$

$$r_{42} = \varepsilon_{13} + a_{23}\varepsilon_{12} + a_{53}\left(\rho_{12} + \frac{i\gamma}{\omega}\right) = \varepsilon_{13} - \frac{\varepsilon_{23}}{\varepsilon_{22}}\varepsilon_{12} + \frac{1}{\mu_{22}}\frac{k_x}{\omega}\frac{i\gamma}{\omega} \tag{73}$$

$$r_{43} = \rho_{11} + a_{24}\varepsilon_{12} + a_{54}\left(\rho_{12} + \frac{i\gamma}{\omega}\right) = \frac{i\gamma}{\omega}\frac{\varepsilon_{12}}{\varepsilon_{22}} - \frac{\mu_{21}}{\mu_{22}}\frac{i\gamma}{\omega} = \frac{i\gamma}{\omega}\left(\frac{\varepsilon_{12}}{\varepsilon_{22}} - \frac{\mu_{21}}{\mu_{22}}\right) \tag{74}$$

$$r_{44} = \rho_{13} + a_{26}\varepsilon_{12} + a_{56}\left(\rho_{12} + \frac{i\gamma}{\omega}\right) = -\frac{k_x}{\omega}\frac{\varepsilon_{12}}{\varepsilon_{22}} - \frac{\mu_{23}}{\mu_{22}}\frac{i\gamma}{\omega} \tag{75}$$

For biaxial electric and magnetic crystalline properties in a principal axis system, only the diagonal elements of the sub tensors of M survive again [see $a'_{ij}$ in (47) and (48)], making (60) — (75) become

$$r_{11} = 0 \;;\; r_{12} = 0 \;;\; r_{13} = \frac{1}{\varepsilon_{22}}\frac{i\gamma}{\omega}\frac{k_x}{\omega} \;;\; r_{14} = \mu_{33} - \frac{1}{\varepsilon_{22}}\frac{k_x}{\omega}\frac{k_x}{\omega} \tag{76}$$

$$r_{21} = 0 \;;\; r_{22} = 0 \;;\; r_{23} = -\mu_{11} + \frac{1}{\varepsilon_{22}}\frac{i\gamma}{\omega}\frac{i\gamma}{\omega} \;;\; r_{24} = -\frac{1}{\varepsilon_{22}}\frac{k_x}{\omega}\frac{i\gamma}{\omega} \tag{77}$$

$$r_{31} = -\frac{1}{\mu_{22}}\frac{i\gamma}{\omega}\frac{k_x}{\omega} \;;\; r_{32} = -\varepsilon_{33} + \frac{1}{\mu_{22}}\frac{k_x}{\omega}\frac{k_x}{\omega} \;;\; r_{33} = 0 \;;\; r_{34} = 0 \tag{78}$$

$$r_{41} = \varepsilon_{11} - \frac{1}{\mu_{22}}\frac{i\gamma}{\omega}\frac{i\gamma}{\omega} \;;\; r_{42} = \frac{1}{\mu_{22}}\frac{k_x}{\omega}\frac{i\gamma}{\omega} \;;\; r_{43} = 0 \;;\; r_{44} = 0 \tag{79}$$

The Green s function problem is posed by placing a Dirac delta forcing function

$$\mathbf{J}_{s\delta}(x) = (\hat{x} + \hat{z})\delta(x - x') \tag{80}$$

on the strip conductor (could be an ordinary metal, a low temperature superconductor, a medium temperature MgB$_2$ superconductor, or a ceramic perovskite high temperature superconductor HTSC, for example) and solving the partial differential equation system in space subject to appropriate boundary and interfacial conditions. Figure 1 shows an example structure with two layers, one interface, and one strip conductor (this particular structure will be numerically studied later in Section IV). Equation (80) says that a surface current of unit delta magnitude is impressed in the x- and z- directions. This is consistent with the strip having width w in the x-direction, infinitesimal extent in the y-direction, and extending infinitely in the z-direction corresponding to a uniform cross-section. Because we have Fourier transformed the problem into the spectral domain, the impressed delta current now appears as

$$\mathbf{J}_{s\delta}(k_x) = (\hat{x}+\hat{z})\int_{-b}^{b}\delta(x-x')e^{-ik_x x'}dx' = (\hat{x}+\hat{z})e^{-ik_x x'} \qquad (81)$$

where 2b is the finite width of the enclosure bounding the x extent. Of course, the actual current is a continuous superposition of weighted contributions over the strip width,

$$\mathbf{J}_s(x) = \int_{-w/2}^{w/2}[J_{sx}(x')\hat{x}+J_{sz}(x')\hat{z}]\delta(x-x')dx = \int_{-w/2}^{w/2}\mathbf{J}_s(x')\delta(x-x')dx \qquad (82)$$

Here we have used the fact that current exists only on the strip. Equation (82) merely states that scanning the extent of the strip (with the delta function) will reproduce the correct current distribution function. Now one can state that (3) and (19) having assumed a time-harmonic variation, or (10) and (20) assuming a z-directed propagation constant also, form a complete set of partial differential equations subject to the interfacial conditions

$$\hat{y}\times(\mathbf{H}^+ - \mathbf{H}^-) = \mathbf{J}_s(x) \qquad (83)$$
$$\mathbf{E}_t^+ = \mathbf{E}_t^- \qquad (84)$$

and boundary conditions

$$E_y(x,y) = E_z(x,y) = 0 \; ; \; x = \pm b \qquad (85)$$
$$E_x(x,y) = E_z(x,y) = 0 \; ; \; y = 0, h_T \qquad (86)$$

Equation (83), which arose from curl equation (3), says that the tangential **H** field above the interface minus that below is related to the surface current at that interface. If we take this interface to be where there are conductor strips, $\mathbf{J}_s(x) › 0$, but at other interfaces without strips, $\mathbf{J}_s(x) = 0$ and tangential **H** field continuity occurs. Equation (84) assures tangential electric field **E** continuity at any interface. Equation (85) enables the finite Fourier transform, and (86) constrains the device to be fully enclosed with actual (or computational) walls, where $h_T$ is the total vertical structure thickness.

In the Green's function construction, (80) is impressed on the system through (83) which creates the field solution

$$\overline{\overline{\mathbf{G}}}(x,y;x') = F_L[\delta(x-x')] \qquad (87)$$

Here the system linear operator $F_L$ takes the delta function $\delta(x-x')$ applied in either the $\hat{x}$ or $\hat{z}$ directions and determines the field component responses, making a two indexed tensor (dyadic) of size 6×2. Multiply (87) on the right by $\mathbf{J}_s(x')$ and integrate, and because $F_L$ is a linear operator, the current vector along with the integral operator may be pulled inside it, giving

$$\int_{-b}^{b} \overline{\overline{\mathbf{G}}}(x,y;\ x') \bullet \mathbf{J}_s(x')dx' \ = \ F_L\left[\int_{-b}^{b} \mathbf{J}_s(x')\delta(x-x')dx'\right] \tag{88}$$

The left-hand side is the field solution of the problem $\mathbf{F}(x,y)$, and because the argument of the linear operator by (82) is the total vector surface current, (88) yields

$$\mathbf{F}(x,y) \ = \ F_L[\mathbf{J}_s(x)] \tag{89}$$

Therefore, with knowledge of $\overline{\overline{\mathbf{G}}}(x,y;\ x')$, the field solution is immediately known,

$$\mathbf{F}(x,y) \ = \ \int_{-b}^{b} \overline{\overline{\mathbf{G}}}(x,y;\ x') \bullet \mathbf{J}_s(x')dx' \tag{90}$$

Considering $\mathbf{J}_s$ as a form of a field, as well as $\mathbf{F}$ being a field, makes (90) an integral equation of the homogeneous Fredholm type of the second kind [15]. Neither $\mathbf{J}_s$ nor $\mathbf{F}$ are known. They must be found by solving (90), with the understanding that the kernel $\overline{\overline{\mathbf{G}}}(x,y;\ x')$ can be acquired before finding the unknowns. Because we will be working in the spectral domain, the integral equation of the problem (90) must be converted to this domain also. Before we do this, note that if the delta function sources were anywhere in the cross-section, it would be written as $\delta(\underline{\rho}-\underline{\rho}')$, implying then by extension of (87) with it being operated on by $F_L$, that the spatial Green's function will also be a function of $\underline{\rho}-\underline{\rho}'$, where $\underline{\rho} = x\hat{x} + y\hat{y}$. That is,

$$\overline{\overline{\mathbf{G}}}(\underline{\rho}-\underline{\rho}') \ = \ F_L[\delta(\underline{\rho}-\underline{\rho}')] \tag{91}$$

Equation (90) would then be written as

$$\mathbf{F}(x,y) \ = \ \int_{'b}^{b} dx' \int_{0}^{h_T} dy' \overline{\overline{\mathbf{G}}}(\underline{\rho}-\underline{\rho}') \bullet \mathbf{J}_s(\underline{\rho}') \tag{92}$$

This integral equation is very general and can be solved for any number of interfacial layers with conductor strips. However, because later in the paper we are restricting ourselves to the guiding of waves in a multi-layered structure with one interface containing a guiding conductor, it is only the form of (92) which is instructive — it is a multi-dimensional convolution integral. In (92) the differential integration element $\rho d\theta d\rho$ was not used in order to explicitly indicate the cross-sectional dimensions. Clearly, as before, the kernel $\overline{\overline{\mathbf{G}}}(\underline{\rho}-\underline{\rho}')$ can be acquired, and (92) solved for the unknown fields. However, there are several ways to find the entire field solution, and one effective way, which does not require obtaining the dyadic Green s function over the whole cross-sectional spatial domain, uses the fact that the strip is a perfect conductor with

$$\mathbf{E}_t(x,y) = 0 \ : \ |x| < w \ , \ y = y_I \tag{93}$$

Here w is the physical width of the strip, and $y_I$ its location along the y axis. A smaller piece of the $\overline{\overline{\mathbf{G}}}$ must be used, $\overline{\overline{\mathbf{G}}}_{EJ}^{xz}$, which relates the driving surface current to the two tangential electric field components. At the $y = y_I$ interface, (92) is cast into the form

$$\mathbf{E}(x,y_I) = \int_{-b}^{b} \overline{\overline{\mathbf{G}}}_{EJ}^{xz}(x-x';y_I) \bullet \mathbf{J}_s(x')dx' \tag{94}$$

Forms (92) and (94) of the integral equation are convolution integrals of the kernel and the driving surface current. They both have the wonderful property that transformation into the spectral domain for respectively 2D and 1D removes the integral operation. The solution procedure employed requires us to take a finite Fourier transform of (94),

$$\begin{aligned}
\int_{-b}^{b} \mathbf{E}(x)e^{-ik_x x}dx &= \int_{-b}^{b} e^{-ik_x x}dx \left\{ \int_{-b}^{b} \overline{\overline{\mathbf{G}}}_{EJ}^{xz}(x-x') \bullet \mathbf{J}_s(x')dx' \right\} \\
&= \int_{-b}^{b} \left\{ \int_{-b}^{b} \overline{\overline{\mathbf{G}}}_{EJ}^{xz}(x-x')e^{-ik_x x}dx \right\} \bullet \mathbf{J}_s(x')dx' \\
&= \int_{-b}^{b} \left\{ \int_{-b}^{b} \overline{\overline{\mathbf{G}}}_{EJ}^{xz}(x'')e^{-ik_x(x'+x'')}dx'' \right\} \bullet \mathbf{J}_s(x')dx' \\
&= \int_{-b}^{b} \left\{ \int_{-b}^{b} \overline{\overline{\mathbf{G}}}_{EJ}^{xz}(x'')e^{-ik_x x''}dx'' \right\} \bullet \mathbf{J}_s(x')e^{-ik_x x'}dx' \\
&= \left\{ \int_{-b}^{b} \overline{\overline{\mathbf{G}}}_{EJ}^{xz}(x'')e^{-ik_x x''}dx'' \right\} \bullet \left\{ \int_{-b}^{b} \mathbf{J}_s(x')e^{-ik_x x'}dx' \right\} \\
&= \overline{\overline{\mathbf{G}}}_{EJ}^{xz}(k_x) \bullet \mathbf{J}_s(k_x)
\end{aligned} \tag{95}$$

or

$$\mathbf{E}(k_x) = \overline{\overline{\mathbf{G}}}_{EJ}^{xz}(k_x) \bullet \mathbf{J}_s(k_x) \tag{96}$$

In (95) $x'' = x - x'$, $dx'' = dx$, led to the third step, and (13) to the final step. $\overline{\overline{\mathbf{G}}}_{EJ}^{xz}$ is given by [16]

$$\tilde{G}_{xx}(k_x,\gamma) = \frac{P_{13}^{(1)}\left[P_{24}^{(21)}P_{14}^{(2)} - P_{14}^{(21)}P_{24}^{(2)}\right]}{P_{14}^{(21)}P_{23}^{(21)} - P_{13}^{(21)}P_{24}^{(21)}} + \frac{P_{14}^{(1)}\left[P_{13}^{(21)}P_{24}^{(2)} - P_{23}^{(21)}P_{14}^{(2)}\right]}{P_{14}^{(21)}P_{23}^{(21)} - P_{13}^{(21)}P_{24}^{(21)}} \tag{97}$$

$$\tilde{G}_{xz}(k_x,\gamma) = -\frac{P_{13}^{(1)}\left[P_{24}^{(21)}P_{13}^{(2)} - P_{14}^{(21)}P_{23}^{(2)}\right]}{P_{14}^{(21)}P_{23}^{(21)} - P_{13}^{(21)}P_{24}^{(21)}} - \frac{P_{14}^{(1)}\left[P_{13}^{(21)}P_{23}^{(2)} - P_{23}^{(21)}P_{13}^{(2)}\right]}{P_{14}^{(21)}P_{23}^{(21)} - P_{13}^{(21)}P_{24}^{(21)}} \tag{98}$$

$$\tilde{G}_{zx}(k_x,\gamma) = \frac{P_{23}^{(1)}\left[P_{24}^{(21)}P_{14}^{(2)} - P_{14}^{(21)}P_{24}^{(2)}\right]}{P_{14}^{(21)}P_{23}^{(21)} - P_{13}^{(21)}P_{24}^{(21)}} + \frac{P_{24}^{(1)}\left[P_{13}^{(21)}P_{24}^{(2)} - P_{23}^{(21)}P_{14}^{(2)}\right]}{P_{14}^{(21)}P_{23}^{(21)} - P_{13}^{(21)}P_{24}^{(21)}} \tag{99}$$

$$\tilde{G}_{zz}(k_x,\gamma) = -\frac{P_{23}^{(1)}\left[P_{24}^{(21)}P_{13}^{(2)} - P_{14}^{(21)}P_{23}^{(2)}\right]}{P_{14}^{(21)}P_{23}^{(21)} - P_{13}^{(21)}P_{24}^{(21)}} - \frac{P_{24}^{(1)}\left[P_{13}^{(21)}P_{23}^{(2)} - P_{23}^{(21)}P_{13}^{(2)}\right]}{P_{14}^{(21)}P_{23}^{(21)} - P_{13}^{(21)}P_{24}^{(21)}} \tag{100}$$

with

$$P^{(1)} = P^{(1)}(h_1) \ ; \ P^{(2)} = P^{(2)}(h_2) \ ; \ P^{(21)} = P^{(2)}(h_2)P^{(1)}(h_1) \quad (101)$$

$$P(y) = e^{i\omega \mathbf{R} y} \quad (102)$$

where $h_i$ = thickness of the $i$ th layer. Note that $\mathbf{R}$ is the same matrix operator first introduced in (53), governing the field behavior variation in the y direction. It contains both the physical properties of the LHM (or RHM) and the electromagnetic field equations.

III. DETERMINATION OF THE EIGENVALUES AND EIGENVECTORS for LHM or DNM

Formulas (97) — (100) were found utilizing (53) which has a solution in the $i$th layer of

$$\aleph_i = \aleph(y_i) = P^{(i)}(y_i)\aleph_i(h_{Ti}) \quad (103)$$

using the global y coordinate. It is very convenient to convert to the local coordinates $y'_i$ where for the $i$ th layer now the local coordinate ($h_0 \equiv 0$) is

$$y'_i = y_i - h_{Ti} \ ; \ h_{Ti} = \sum_{k=0}^{i-1} h_k \quad (104)$$

with $h_{Ti}$ being the total thickness of all the layers prior to the $i$ th layer. In the local coordinate system, (103) appears as

$$\aleph_i = \aleph(y'_i) = P^{(i)}(y'_i)\aleph_i(0) \quad (105)$$

This equation is applied repeatedly to each layer throughout the structure, being careful to impose (83) — (86) in the spectral domain:

$$\hat{y} \times \left[\mathbf{H}^+(k_x,y) - \mathbf{H}^-(k_x,y)\right] = \mathbf{J}_s(k_x) \ ; \ y = y_I \quad (106)$$

$$\mathbf{E}_t^+(k_x,y) = \mathbf{E}_t^-(k_x,y) \ ; \ y = h_i + h_{Ti} = \sum_{k=0}^{i} h_k \quad (107)$$

and boundary conditions

$$E_x(k_x,y) = E_z(k_x,y) = 0 \ ; \ y = 0, h_T \quad (108)$$

Boundary condition on the side walls (85) get converted to the spectral domain in a process which generates the discretization of $k_x$. The detailed derivation will be given since the whole technique hinges on it.

By (14), the spatial electric field components $E_{y,z}$ are expressed as

$$E_{y,z}(x,y) = \frac{1}{2b} \sum_{k_x = -\infty}^{\infty} E_{y,z}(k_x,y)e^{ik_x x} \quad (109)$$

Consider first the case where $E_{y,z}(x,y)$ has even symmetry with respect to the x-axis. (Symmetry choices will be covered in more detail after the derivation of the discretization $k_x$ values.) For even symmetry,

$$E_{y,z}(x,y) = E_{y,z}(-x,y) \tag{110}$$

Invoking (109), this becomes

$$\frac{1}{2b}\sum_{k_x=-\infty}^{\infty}E_{y,z}(k_x,y)e^{ik_xx} = \frac{1}{2b}\sum_{k_x=-\infty}^{\infty}E_{y,z}(k_x,y)e^{-ik_xx}$$
$$= \frac{1}{2b}\sum_{k_x=+\infty}^{-\infty}E_{y,z}(-k_x,y)e^{ik_xx} \tag{111}$$

$$\frac{1}{2b}\sum_{k_x=-\infty}^{\infty}\left[E_{y,z}(k_x,y) - E_{y,z}(-k_x,y)\right]e^{ik_xx} = 0 \tag{112}$$

Equation (112) is true for any x if the bracketed term is zero, namely that

$$E_{y,z}(k_x,y) = E_{y,z}(-k_x,y) \tag{113}$$

Now we must insert this back into the expansion (109), obtaining

$$E_{y,z}(x,y) = \frac{1}{2b}\sum_{k_x=-\infty}^{\infty}E_{y,z}(k_x,y)e^{ik_xx}$$
$$= \frac{1}{2b}\sum_{k_x=-\infty}^{0^-}E_{y,z}(k_x,y)e^{ik_xx} + E_{y,z}(0,y) + \frac{1}{2b}\sum_{k_x=0^+}^{\infty}E_{y,z}(k_x,y)e^{ik_xx}$$
$$= \frac{1}{2b}\sum_{k_x=+\infty}^{0^+}E_{y,z}(-k_x,y)e^{-ik_xx} + E_{y,z}(0,y) + \frac{1}{2b}\sum_{k_x=0^+}^{\infty}E_{y,z}(k_x,y)e^{ik_xx}$$
$$= \frac{1}{2b}\sum_{k_x=0^+}^{\infty}E_{y,z}(k_x,y)e^{-ik_xx} + E_{y,z}(0,y) + \frac{1}{2b}\sum_{k_x=0^+}^{\infty}E_{y,z}(k_x,y)e^{ik_xx} \tag{114}$$
$$= \frac{1}{2b}\sum_{k_x=0^+}^{\infty}E_{y,z}(k_x,y)\left[e^{-ik_xx}+e^{ik_xx}\right] + E_{y,z}(0,y)$$
$$= \frac{1}{b}\sum_{k_x=0^+}^{\infty}E_{y,z}(k_x,y)\cos(k_xx) + E_{y,z}(0,y)$$

In (114), (113) was used for the third step. Imposition of boundary condition (85) forces (114) to obey

$$\frac{1}{b}\sum_{k_x=0^+}^{\infty}E_{y,z}(k_x,y)\cos(k_xx) + E_{y,z}(0,y) = 0 \tag{115}$$

or

$$\cos(k_xb) = 0 \quad ; \quad E_{y,z}(0,y) = 0 \tag{116}$$

The first constraint in (116) restricts $k_x$ to

$$k_x = \frac{2n-1}{2b}\pi \; ; \; n = 0, \pm 1, \pm 2, \cdots \qquad (117)$$

showing that $k_x > 0$, allowing us to drop the second (116) constraint. For odd symmetry,

$$E_{y,z}(x,y) = -E_{y,z}(-x,y) \qquad (118)$$

Invoking (109), this becomes

$$\frac{1}{2b}\sum_{k_x=-\infty}^{\infty}E_{y,z}(k_x,y)e^{ik_x x} = -\frac{1}{2b}\sum_{k_x=-\infty}^{\infty}E_{y,z}(k_x,y)e^{-ik_x x}$$
$$= -\frac{1}{2b}\sum_{k_x=+\infty}^{-\infty}E_{y,z}(-k_x,y)e^{ik_x x} \qquad (119)$$

$$\frac{1}{2b}\sum_{k_x=-\infty}^{\infty}\left[E_{y,z}(k_x,y)+E_{y,z}(-k_x,y)\right]e^{ik_x x} = 0 \qquad (120)$$

Equation (112) is true for any x if the bracketed term is zero, namely that

$$E_{y,z}(k_x,y) = -E_{y,z}(-k_x,y) \qquad (121)$$

Now we must insert this back into the expansion (109), obtaining

$$E_{y,z}(x,y) = \frac{1}{2b}\sum_{k_x=-\infty}^{\infty}E_{y,z}(k_x,y)e^{ik_x x}$$
$$= \frac{1}{2b}\sum_{k_x=-\infty}^{0^-}E_{y,z}(k_x,y)e^{ik_x x} + E_{y,z}(0,y) + \frac{1}{2b}\sum_{k_x=0^+}^{\infty}E_{y,z}(k_x,y)e^{ik_x x}$$
$$= \frac{1}{2b}\sum_{k_x=+\infty}^{0^+}E_{y,z}(-k_x,y)e^{-ik_x x} + E_{y,z}(0,y) + \frac{1}{2b}\sum_{k_x=0^+}^{\infty}E_{y,z}(k_x,y)e^{ik_x x} \qquad (122)$$
$$= -\frac{1}{2b}\sum_{k_x=0^+}^{\infty}E_{y,z}(k_x,y)e^{-ik_x x} + E_{y,z}(0,y) + \frac{1}{2b}\sum_{k_x=0^+}^{\infty}E_{y,z}(k_x,y)e^{ik_x x}$$
$$= \frac{1}{2b}\sum_{k_x=0^+}^{\infty}E_{y,z}(k_x,y)\left[-e^{-ik_x x}+e^{ik_x x}\right] + E_{y,z}(0,y)$$
$$= \frac{i}{b}\sum_{k_x=0^+}^{\infty}E_{y,z}(k_x,y)\sin(k_x x) + E_{y,z}(0,y)$$

In (122), (121) was used in the third step. Imposition of boundary condition (85) forces (122) to obey

$$\pm \frac{i}{b} \sum_{k_x = 0^+}^{\infty} E_{y,z}(k_x, y)\sin(k_x b) + E_{y,z}(0, y) = 0 \tag{123}$$

or

$$\sin(k_x b) = 0 \quad ; \quad E_{y,z}(0, y) = 0 \tag{124}$$

The first constraint in (124) restricts $k_x$ to

$$k_x = \frac{n}{b}\pi \quad ; \quad n = 0, \pm 1, \pm 2, \cdots \tag{125}$$

Since the first constraint allows $k_x = 0$, technically the first summation in (123) does not have n = 0 in its domain, but by widening its domain to cover $k_x = 0$, the second constraint may be dropped. That is, (123) becomes

$$\pm \frac{i}{b} \sum_{k_x = 0}^{\infty} E_{y,z}(k_x, y)\sin(k_x b) = 0 \tag{126}$$

and rule (125) is exact.

Next we treat the origin of the symmetry choices. Go to the harmonic equations (3), using $\nabla \times \mathbf{F} = \varepsilon^{ijk}\nabla_i E_j \hat{x}_k$ to expand them out by components for the doubly biaxial case (biaxial for $\overline{\overline{\varepsilon}}$ and $\overline{\overline{\mu}}$ with principal axes in the coordinate directions), find

$$\frac{\partial H_z}{\partial y} - \frac{\partial H_y}{\partial z} = i\omega\varepsilon_{xx}E_x + J_x \; ; \; \frac{\partial H_x}{\partial z} - \frac{\partial H_z}{\partial x} = i\omega\varepsilon_{yy}E_y \; ; \; \frac{\partial H_y}{\partial x} - \frac{\partial H_x}{\partial y} = i\omega\varepsilon_{zz}E_z + J_z \tag{127}$$

$$\frac{\partial E_z}{\partial y} - \frac{\partial E_y}{\partial z} = -i\omega\mu_{xx}H_x \; ; \; \frac{\partial E_x}{\partial z} - \frac{\partial E_z}{\partial x} = -i\omega\mu_{yy}H_y \; ; \; \frac{\partial E_y}{\partial x} - \frac{\partial E_x}{\partial y} = -i\omega\mu_{zz}H_z \tag{128}$$

Setting $J_z(x,y)$ = even for the impressed current, we find that (127) requires that $E_z(x,y)$ = even, $H_y(x,y)$ = odd and $H_x(x,y)$ = even in its third equation; $H_z(x,y)$ = odd and $E_y(x,y)$ = even in its second equation; and $E_x(x,y)$ = odd and $J_x(x,y)$ = odd in its first equation. These selections are consistent with (128). For $J_z(x,y)$ = odd, all of the selections are reversed. To see that great care must be exercised in this process, look at the ferrite spin system which for principal axes (defined by three orthogonal bias fields) in the three coordinate directions, $\overline{\overline{\mu}}$ is given by [17] (see also [18] for a magnetized semiconductor with similar permittivity tensor)

$$\overline{\overline{\mu}}(H_0\hat{x}) = \begin{bmatrix} \mu_e & 0 & 0 \\ 0 & \mu & -i\kappa \\ 0 & i\kappa & \mu \end{bmatrix} \; ; \; \overline{\overline{\mu}}(H_0\hat{y}) = \begin{bmatrix} \mu & 0 & i\kappa \\ 0 & \mu_e & 0 \\ -i\kappa & 0 & \mu \end{bmatrix} \; ; \; \overline{\overline{\mu}}(H_0\hat{z}) = \begin{bmatrix} \mu & -i\kappa & 0 \\ i\kappa & \mu & 0 \\ 0 & 0 & \mu_e \end{bmatrix} \tag{129}$$

Now for (3) with the static bias field $H_0\hat{x}$ in (129), (127) is still valid for the permittivity biaxial, but (128) is changed to

$$\frac{\partial E_z}{\partial y} - \frac{\partial E_y}{\partial z} = -i\omega\mu_e H_x \ ; \ \frac{\partial E_x}{\partial z} - \frac{\partial E_z}{\partial x} = -i\omega\mu H_y - \kappa H_z \ ; \ \frac{\partial E_z}{\partial y} - \frac{\partial E_y}{\partial z} = -i\omega\mu H_z - \kappa H_y \quad (130)$$

Again imposing $J_z(x,y)$ = even, the same results are obviously seen from (127) for the electrically biaxial ferrite case, with these selections being now consistent also with (130). With the bias field being $H_0\,\hat{y}$, (128) changes to

$$\frac{\partial E_z}{\partial y} - \frac{\partial E_y}{\partial z} = -i\omega\mu H_x - \kappa H_z \ ; \ \frac{\partial E_x}{\partial z} - \frac{\partial E_z}{\partial x} = -i\omega\mu_e H_y \ ; \ \frac{\partial E_z}{\partial y} - \frac{\partial E_y}{\partial z} = -i\omega\mu H_z - \kappa H_x \quad (131)$$

Utilizing the choices found from (127) for $J_z(x,y)$ = even, makes (131) symmetry wise inconsistent, as can be easily verified by inspection. This means that for the $H_0\,\hat{y}$ bias case, all of the fields must be a superposition of both symmetries. Finally for the third tensor permeability in (129), (128) changes to

$$\frac{\partial E_z}{\partial y} - \frac{\partial E_y}{\partial z} = -i\omega\mu H_x - \kappa H_y \ ; \ \frac{\partial E_x}{\partial z} - \frac{\partial E_z}{\partial x} = -i\omega\mu H_y - \kappa H_x \ ; \ \frac{\partial E_z}{\partial y} - \frac{\partial E_y}{\partial z} = -i\omega\mu_e H_z \quad (132)$$

This last case for $H_0\,\hat{z}$ bias also has the symmetry inconsistency between (127) and curl equation (132) with the magnetic constitutive information, requiring all fields to be a superposition of both symmetry types.

    The surface currents which drive the problem self-consistently, can be chosen in a number of ways, it only being necessary to prepare complete sets of basis functions which are used to construct them. They are selected in the real space domain to display some advantageous property, for example, edge singularity behavior due to charge repulsion. For the complete set of cosinusoidal basis functions modified by the edge condition, we have for a strip with even mode symmetry (determined by the z-component symmetry as just discussed above)

$$J_{zm}(x) = \xi_{em}(x) = \begin{cases} \dfrac{\cos\left(\pi \dfrac{x}{w}[m-1]\right)}{\sqrt{1-(x/w)^2}} & ; \ |x| \leq w \\ 0 & ; \ w < |x| \end{cases} \quad (133)$$

$$J_{xm}(x) = \eta_{em}(x) = \begin{cases} \dfrac{\sin\left(\pi \dfrac{x}{w} m\right)}{\sqrt{1-(x/w)^2}} & ; \ |x| \leq w \\ 0 & ; \ w < |x| \end{cases} \quad (134)$$

and for odd mode symmetry,

$$J_{zm}(x) = \xi_{om}(x) = \begin{cases} \dfrac{\sin\left(\dfrac{\pi}{2} \dfrac{x}{w}[2m-1]\right)}{\sqrt{1-(x/w)^2}} & ; \ |x| \leq w \\ 0 & ; \ w < |x| \end{cases} \quad (135)$$

$$J_{xm}(x) = \eta_{om}(x) = \begin{cases} \dfrac{\cos\left(\dfrac{\pi}{2}\dfrac{x}{w}[2m-1]\right)}{\sqrt{1-(x/w)^2}} & ; \quad |x| \le w \\ 0 & ; \quad w < |x| \end{cases} \quad (136)$$

Superposition of the complete set forms the total surface current (and constitutes the moment method when unknown currents/fields are expanded using basis sets and then inner products with weights are then taken on the structure governing equation to derive a linear system to be solved)

$$J_{xe}(x) = \sum_{m=1}^{n_x} a_{em}\eta_{em}(x) \quad ; \quad J_{ze}(x) = \sum_{m=1}^{n_z} b_{em}\xi_{em}(x) \quad (137)$$

$$J_{xo}(x) = \sum_{m=1}^{n_x} a_{om}\eta_{om}(x) \quad ; \quad J_{zo}(x) = \sum_{m=1}^{n_z} b_{om}\xi_{om}(x) \quad (138)$$

Fourier transforming (137) and (138) according to (13) gives

$$J_{xe}(n) = \sum_{m=1}^{n_x} a_{em}\eta_{em}(n) \quad ; \quad J_{ze}(n) = \sum_{m=1}^{n_z} b_{em}\xi_{em}(n) \quad (139)$$

$$J_{xo}(n) = \sum_{m=1}^{n_x} a_{om}\eta_{om}(n) \quad ; \quad J_{zo}(n) = \sum_{m=1}^{n_z} b_{om}\xi_{om}(n) \quad (140)$$

with

$$\xi_{em}(n) = \xi_{em}(k_x[n]) = \frac{\pi w}{2}\{J_0(k_x w + [m-1]\pi) + J_0(k_x w - [m-1]\pi)\} \quad (141)$$

$$\eta_{em}(n) = \eta_{em}(k_x[n]) = -\frac{i\pi w}{2}\{J_0(k_x w + m\pi) - J_0(k_x w - m\pi)\} \quad (142)$$

$$\xi_{om}(n) = \xi_{om}(k_x[n]) = -\frac{i\pi w}{2}\left\{J_0\left(k_x w + [2m-1]\frac{\pi}{2}\right) - J_0\left(k_x w - [2m-1]\frac{\pi}{2}\right)\right\} \quad (143)$$

$$\eta_{om}(n) = \eta_{om}(k_x[n]) = \frac{\pi w}{2}\left\{J_0\left(k_x w + [2m-1]\frac{\pi}{2}\right) + J_0\left(k_x w - [2m-1]\frac{\pi}{2}\right)\right\} \quad (144)$$

An exact solution is obtained only when $n_x$ and $n_z \to$ . However, a finite number of them may be used, depending on the propagating eigenmode modeled, to find a reasonably accurate numerical result. In (141) — (144), $J_0$ denotes the Bessel function of the first kind.

Eigenvalues $\gamma$ and eigenvectors of the propagating problem can now be found from (96), the interfacial strip equation (drop all interfacial indexes, and use subscripts to label elements),

$$\begin{aligned} E_x(n,\gamma) &= G_{xx}(\gamma,n)J_x(n,\gamma) + G_{xz}(\gamma,n)J_z(n,\gamma) \\ E_z(n,\gamma) &= G_{zx}(\gamma,n)J_x(n,\gamma) + G_{zz}(\gamma,n)J_z(n,\gamma) \end{aligned} \quad (145)$$

by substituting for the surface currents using expressions (139) or (140) depending upon the mode symmetry to be studied (we drop the explicit mode symmetry type notation since we will here only treat one or the other type of pure symmetry solution).

$$E_x(n,\gamma) = G_{xx}(\gamma,n)\sum_{i=1}^{n_x} a_i\eta_i(n) + G_{xz}(\gamma,n)\sum_{i=1}^{n_z} b_i\xi_i(n)$$

$$E_z(n,\gamma) = G_{zx}(\gamma,n)\sum_{i=1}^{n_x} a_i\eta_i(n) + G_{zz}(\gamma,n)\sum_{i=1}^{n_z} b_i\xi_i(n) \quad (146)$$

Next multiply the first equation of (146) by $\eta_j$ and the second by $\xi_j$, then summing over the spectral index n (this is the inner product part of the moment method)

$$\sum_{n=-\infty}^{\infty}\eta_j(n)E_x(n,\gamma) = \sum_{n=-\infty}^{\infty}\left[\eta_j(n)G_{xx}(\gamma,n)\sum_{i=1}^{n_x} a_i(n)\eta_i(n)\right] + \sum_{n=-\infty}^{\infty}\left[\eta_j(n)G_{xz}(\gamma,n)\sum_{i=1}^{n_z} b_i(n)\xi_i(n)\right]$$

$$\sum_{n=-\infty}^{\infty}\xi_j(n)E_z(n,\gamma) = \sum_{n=-\infty}^{\infty}\left[\xi_j(n)G_{zx}(\gamma,n)\sum_{i=1}^{n_x} a_i(n)\eta_i(n)\right] + \sum_{n=-\infty}^{\infty}\left[\xi_j(n)G_{zz}(\gamma,n)\sum_{i=1}^{n_z} b_i(n)\xi_i(n)\right] \quad (147)$$

Interchanging the order of the basis function and spectral summations in (147),

$$\sum_{n=-\infty}^{\infty}\eta_j(n)E_x(n,\gamma) = \sum_{i=1}^{n_x} a_i(n)\left[\sum_{n=-\infty}^{\infty}\eta_j(n)G_{xx}(\gamma,n)\eta_i(n)\right] + \sum_{i=1}^{n_z} b_i(n)\left[\sum_{n=-\infty}^{\infty}\eta_j(n)G_{xz}(\gamma,n)\xi_i(n)\right]$$

$$\sum_{n=-\infty}^{\infty}\xi_j(n)E_z(n,\gamma) = \sum_{i=1}^{n_x} a_i(n)\left[\sum_{n=-\infty}^{\infty}\xi_j(n)G_{zx}(\gamma,n)\eta_i(n)\right] + \sum_{i=1}^{n_z} b_i(n)\left[\sum_{n=-\infty}^{\infty}\xi_j(n)G_{zz}(\gamma,n)\xi_i(n)\right] \quad (148)$$

Examine the left hand sides of this paired set of equations (148):

$$\begin{aligned}
\begin{bmatrix} -1 & 1 \\ 1 & -1 \end{bmatrix} \sum_{n=-\infty}^{\infty} \begin{Bmatrix} \eta_{e,o;j} \\ \xi_{e,o;j} \end{Bmatrix}(n) E_{x,z}(n,\gamma) &= \sum_{n=-\infty}^{\infty} \begin{Bmatrix} \eta_j^* \\ \xi_j^* \end{Bmatrix}(n) E_{x,z}(n,\gamma) \\
&= \sum_{n=-\infty}^{\infty} \left[ \int_{-b}^{b} \begin{Bmatrix} \eta_j \\ \xi_j \end{Bmatrix}(x) e^{-ik_x x} dx \right]^* E_{x,z}(n,\gamma) \\
&= \sum_{n=-\infty}^{\infty} \int_{-b}^{b} \begin{Bmatrix} \eta_j^* \\ \xi_j^* \end{Bmatrix}(x) e^{ik_x x} dx E_{x,z}(n,\gamma) \\
&= \sum_{n=-\infty}^{\infty} \int_{-b}^{b} \begin{Bmatrix} \eta_j \\ \xi_j \end{Bmatrix}(x) e^{ik_x x} dx E_{x,z}(n,\gamma) \qquad (149) \\
&= \int_{-b}^{b} \begin{Bmatrix} \eta_j \\ \xi_j \end{Bmatrix}(x) \left[ \sum_{n=-\infty}^{\infty} E_{x,z}(n,\gamma) e^{ik_x x} \right] dx \\
&= 2b \int_{-b}^{b} \begin{Bmatrix} \eta_j \\ \xi_j \end{Bmatrix}(x) E_{x,z}(x,\gamma) dx \\
&= 0
\end{aligned}$$

where the first and second rows in the left-hand side matrix corresponds, respectively, to even and odd symmetry. Right hand equalities in the 1 st, 2 nd, 4 th, 6 th and last lines used, respectively, (141) — (144), (13), (133) — (136), (14), and (92) with (133) — (136). Equation (149) amounts to a Parseval theorem [19] for the problem at hand. Enlisting this theorem, (148) can be rewritten as

$$\sum_{i=1}^{n_x} a_i(n) \left[ \sum_{n=-\infty}^{\infty} \eta_j(n) G_{xx}(\gamma,n) \eta_i(n) \right] + \sum_{i=1}^{n_z} b_i(n) \left[ \sum_{n=-\infty}^{\infty} \eta_j(n) G_{xz}(\gamma,n) \xi_i(n) \right] = 0$$

$$\sum_{i=1}^{n_x} a_i(n) \left[ \sum_{n=-\infty}^{\infty} \xi_j(n) G_{zx}(\gamma,n) \eta_i(n) \right] + \sum_{i=1}^{n_z} b_i(n) \left[ \sum_{n=-\infty}^{\infty} \xi_j(n) G_{zz}(\gamma,n) \xi_i(n) \right] = 0 \qquad (150)$$

Since (150) is true for any $j$ th basis test function, it may be condensed into the form

$$\sum_{i=1}^{n_x} a_i(n) X_{xx}^{ji}(\gamma) + \sum_{i=1}^{n_z} b_i(n) X_{xz}^{ji}(\gamma) = 0 \quad ; \quad j = 1, 2, \cdots n_x$$

$$\sum_{i=1}^{n_x} a_i(n) X_{zx}^{ji}(\gamma) + \sum_{i=1}^{n_z} b_i(n) X_{zz}^{ji}(\gamma) = 0 \quad ; \quad j = 1, 2, \cdots n_z \qquad (151)$$

where (this is referred to as the Galerkin technique since $\{\eta_i, \xi_i\} = \{\eta_j, \xi_j\}$)

$$X_{xx}^{ji}(\gamma) = \sum_{n=-\infty}^{\infty} \eta_j(n) G_{xx}(\gamma,n) \eta_i(n)$$

$$X_{xz}^{ji}(\gamma) = \sum_{n=-\infty}^{\infty} \eta_j(n) G_{xz}(\gamma,n) \xi_i(n)$$

$$X_{zx}^{ji}(\gamma) = \sum_{n=-\infty}^{\infty} \xi_j(n) G_{zx}(\gamma,n) \eta_i(n) \tag{152}$$

$$X_{zz}^{ji}(\gamma) = \sum_{n=-\infty}^{\infty} \xi_j(n) G_{zz}(\gamma,n) \xi_i(n)$$

In matrix form, (151) appears as

$$\begin{bmatrix} X_{xx} & X_{xz} \\ X_{zx} & X_{zz} \end{bmatrix} \begin{bmatrix} a \\ b \end{bmatrix} = 0 \quad ; \quad \mathbf{X} \begin{bmatrix} a \\ b \end{bmatrix} = 0 \quad ; \quad \mathbf{X}\mathbf{v} = 0$$

$$\mathbf{X} = \begin{bmatrix} X_{xx} & X_{xz} \\ X_{zx} & X_{zz} \end{bmatrix} \quad ; \quad \mathbf{v} = \begin{bmatrix} a \\ b \end{bmatrix} \tag{153}$$

Once system of equations (153) is solved, vector **v** containing the coefficients needed to construct the surface current, and from them the electromagnetic field, is known, which is the problem eigenvector. The eigenvalue, $\gamma$, is determined from this system also (from the determinant of **X** being set to zero). From (137) and (138), the total vector surface current is obtained. Defining $\alpha_n = k_x(n)$,

$$\mathbf{J}(x,y) = \frac{1}{2b} \sum_{n=-n_{max}}^{n_{max}} \mathbf{J}(n;y) e^{i\alpha_n x} \tag{154}$$

Once the total surface current is available, the total eigenvector field solution follows from $\overline{\overline{\mathbf{G}}}(\rho - \rho\ll)$ in (92) [12],

$$\mathbf{E}(x,y) = \frac{1}{2b} \sum_{n=-n_{max}}^{n_{max}} \mathbf{E}(n;y) e^{i\alpha_n x} \quad ; \quad \mathbf{H}(x,y) = \frac{1}{2b} \sum_{n=-n_{max}}^{n_{max}} \mathbf{H}(n;y) e^{i\alpha_n x} \tag{155}$$

Spectral expansion is truncated at the same maximum number of terms n = $n_{max}$ for all vector components. Basis function summation limits $n_x$ and $n_z$ for the x and z components (m = $m_{max}$) [see (139) and (140) for the surface current expansion] can be truncated at different values. Current and fields are real physical quantities, so they must be converted through

$$\mathbf{J}_p(x,y,z) = \text{Re}\left[\mathbf{J}(x,y)e^{i\omega t - \gamma z}\right] ; \mathbf{E}_p(x,y,z) = \text{Re}\left[\mathbf{E}(x,y)e^{i\omega t - \gamma z}\right] ; \mathbf{H}_p(x,y,z) = \text{Re}\left[\mathbf{H}(x,y)e^{i\omega t - \gamma z}\right] \tag{156}$$

which reasserts both the time and z-dependence down the guiding structure. At a particular z plane, say z = 0, we may drop out the explicit z-dependence. And if we don t wish to watch the time evolution of the harmonic wave, which is sufficient for plotting purposes, we may further set t = 0, and write (156) as

$$\mathbf{J}_p(x,y) = \text{Re}[\mathbf{J}(x,y)] \quad ; \quad \mathbf{E}_p(x,y) = \text{Re}[\mathbf{E}(x,y)] \quad ; \quad \mathbf{H}_p(x,y) = \text{Re}[\mathbf{H}(x,y)] \tag{157}$$

# IV. NUMERICAL CALCULATIONS OF THE ELECTROMAGNETIC FIELD for LHM or DNM

To enable us to assess the effect of varying the anisotropy on the propagation constant and the field patterns, we will study the single biaxial crystal case, which has the anisotropy in the permittivity. The permeability will be left alone, set to a scalar value of u = -2.5 (relative to the free space value). To further simplify the interpretation, the electric crystalline properties will be chosen with principle axis orientation. Thus the material tensors are

$$\bar{\bar{\varepsilon}} = \begin{bmatrix} \varepsilon_{xx} & 0 & 0 \\ 0 & \varepsilon_{yy} & 0 \\ 0 & 0 & \varepsilon_{zz} \end{bmatrix} \quad ; \quad \bar{\bar{\mu}} = \begin{bmatrix} \mu_{xx} & 0 & 0 \\ 0 & \mu_{yy} & 0 \\ 0 & 0 & \mu_{zz} \end{bmatrix} = \mu \begin{bmatrix} 1 & 0 & 0 \\ 0 & 1 & 0 \\ 0 & 0 & 1 \end{bmatrix} \tag{158}$$

The structure to be modelled is a single microstrip guided wave device (Fig. 1), with the wave propagating in the z-direction, with the cross-section of the device being uniform in every xy-planar cut. Substrate material is the LHM (NIM or NPV) with $h_1$ = 5 mm, which rests over a ground plane, above it is a conductor strip of perfect conductivity with insignificant thickness and width w = 5 mm, and then above it is placed an overlayer of RHM with unity permittivity and permeability with $h_2$ = 20 mm (perfect air or vacuum). This stacking is done in the y-direction. Initially we start out with a nominal permittivity value of $\varepsilon_n$ = -2.5 (relative to the free space value) making ($\mu_n$ = -2.5)

$$\bar{\bar{\varepsilon}} = \begin{bmatrix} \varepsilon_n & 0 & 0 \\ 0 & \varepsilon_n & 0 \\ 0 & 0 & \varepsilon_n \end{bmatrix} \quad ; \quad \bar{\bar{\mu}} = \mu_n \begin{bmatrix} 1 & 0 & 0 \\ 0 & 1 & 0 \\ 0 & 0 & 1 \end{bmatrix} \tag{159}$$

For the nominal values of this device, we are looking at the lowest order mode possible. Spectral terms in (152), and in (154) and (155), are summed to n = $n_{max}$ =200, and the basis function limits are set $n_x = n_z = 1$. Further terms in the expansion are not expected to have a large effect on either the current or field distribution because this number of basis functions is enough to capture the current variation for the lowest order mode which is symmetric. Another salutary effect of being able to use so few basis function terms is that unwanted higher order modes would not be available to create the surface current which excites the structure. Nevertheless, for each propagation constant γ eigenvalue found below, other higher order solutions were found (up to six in some situations) to verify that our γ was the desired one. We will pick out three frequencies to study spanning two decades, each frequency within a range. In the 0.1 — 0.3 GHz range, we select 0.2 GHz, which lies within a range possessing imaginary propagation constants with the phase value β (relative to the free space value) varying from 3.5 to 4.5 [nominal case of (159)]. An order of magnitude higher is the next frequency value at 2.0 GHz, which lies within the 0.3 — 10 GHz range, whose propagation constants γ are complex (γ = α + iβ) with α varying in value from 1.7 to 0.35 and β varying over 4.0 to 0.8 [nominal case of (159)]. Finally, the last frequency chosen is yet one more order of magnitude higher at 20 GHz, in the frequency range 10 — 40 GHz, where again γ= iβ giving pure phase behavior with β varying from 2.0 to 2.5 [nominal case of (159)]. None of the γ versus f variations is linear in any of these three ranges.

All results displayed below were done by partitioning the structure into 45 divisions per layer vertically, and 90 divisions horizontally, producing a mesh with grid points having field values calculated at each of them. Field mesh values were then sent through another processing step to produce the magnitude field distributions. Arrow distributions were produced in the second processing stage by attaching arrows to a courser grid point mesh created from the first meshing scheme, which was on the order of $20 \times 20$ in size.

Figs. 2(a) — (d) [in color, linear scale shown below device cross-section] show the electric field magnitude $E = \left[\sum_{i=1}^{3} E_i^2\right]^{1/2}$ at f = 0.2 GHz for

$$\bar{\bar{\varepsilon}}(a) = \begin{bmatrix} \varepsilon_n & 0 & 0 \\ 0 & \varepsilon_n & 0 \\ 0 & 0 & \varepsilon_n \end{bmatrix}; \bar{\bar{\varepsilon}}(b) = \begin{bmatrix} \varepsilon_n & 0 & 0 \\ 0 & 2\varepsilon_n & 0 \\ 0 & 0 & \varepsilon_n \end{bmatrix}; \bar{\bar{\varepsilon}}(c) = \begin{bmatrix} \varepsilon_n & 0 & 0 \\ 0 & 2\varepsilon_n & 0 \\ 0 & 0 & 2\varepsilon_n \end{bmatrix}; \bar{\bar{\varepsilon}}(d) = \begin{bmatrix} 2\varepsilon_n & 0 & 0 \\ 0 & 2\varepsilon_n & 0 \\ 0 & 0 & \varepsilon_n \end{bmatrix}$$

(160)

Notice that the first tensor $\bar{\bar{\varepsilon}}(a)$ case represents isotropy, so that any deviation from the distributions shown for it, for the cases $\bar{\bar{\varepsilon}}(b)$ — $\bar{\bar{\varepsilon}}(d)$, indicate the effect of anisotropy. With mostly phase behavior except for the last case, complex propagation constant is for these four cases $\alpha(a)$, $\beta(a)$ = 0, 3.753; $\alpha(b)$, $\beta(b)$ = $6.261 \times 10^{-4}$, 5.910; $\alpha(c)$, $\beta(c)$ = $8.112 \times 10^{-4}$, 5.960; $\alpha(d)$, $\beta(d)$ = 0.7835, 7.456. Surface current expansion coefficients (complex, x and z components [see (139)]) for the cases are, respectively, $\{a_{e1}, b_{e1}\}$ = $\{(0, -7.859 \times 10^{-4}); (1,0)\}$, $\{(0, -2.159 \times 10^{-3}); (1,0)\}$, $\{(-3.171 \times 10^{-6}, -2.202 \times 10^{-3}), (1,0)\}$, $\{(-8.875 \times 10^{-5}, -1.133 \times 10^{-3}), (1,0)\}$. Very little change in the electric magnitude distribution occurs until we get to case (d), which has a rather noticeable increase in intensity above the interface, and the shape of the distribution changed or enlarged to look like more of a bubble immediately around and on top of the strip. Substantial $\alpha(d)$ indicates lateral wave motion due to a wave attached to the interface and having some surface wave character, or bulk nature, or both, and must correspond to the energy being carried by a large portion of the top region seen in the Fig. 2 (d). (At the end of this section, we will return to this question by examining field cuts.) Figs. 3(a) — (d) [in color] show the magnetic field magnitude $H = \left[\sum_{i=1}^{3} H_i^2\right]^{1/2}$ for (160). Magnetic magnitude distribution is seen to keep the same basic shape, but the intensity of the field is seen to progressively rise just above the interface, and below it beneath the strip and immediately to either side of the strip. Figs. 4 (a) — (d) show electric field arrow plots of $\mathbf{E}_t$, the field in the plane transverse to the z-direction, with the arrow length indicating the cross-sectional magnitude and the orientation the direction [maximum arrow size is shown below the device cross-section — this value is correlated with the field intensity values shown in Figs. 2 and 3 for the same $\bar{\bar{\varepsilon}}$ cases]. Nominal case for $\mathbf{E}_t$ is given in Figs. 4(a) and (b) which provide, respectively, the actual arrow distribution and the scaled distribution, with the scaled plot lifting the small magnitude arrows out of obscurity so one can study their directions. Scaling is done according to a formula using an inverse trigonometric function,

$$E_i = \frac{E_i}{E_t}\left[\left|\tan^{-1}\left(\frac{E_t}{E_{av}}\right)\right| + 0.75\right] \quad ; \quad H_i = \frac{H_i}{H_t}\left[\left|\tan^{-1}\left(\frac{H_t}{H_{av}}\right)\right| + 0.75\right] \tag{161}$$

$$E_t^2 = E_x^2 + E_y^2 \quad ; \quad H_t^2 = H_x^2 + H_y^2 \tag{162}$$

One should be very careful in using scaled plots to understand anything other than direction behavior. Arrow plots of $\mathbf{E}_t$ for cases $\bar{\bar{\varepsilon}}$(b) and $\bar{\bar{\varepsilon}}$(c) look similar to $\bar{\bar{\varepsilon}}$(a), and because it is much harder to resolve subtle trends in arrow plots versus color magnitude distribution plots, as we have just seen, we omit them and go on to the last $\bar{\bar{\varepsilon}}$(d) case of (160). Figs. 4(c) and (d) provide, respectively, the unscaled and scaled distributions of $\mathbf{E}_t$ for this last case $\bar{\bar{\varepsilon}}$(d). Significant change is seen from the $\bar{\bar{\varepsilon}}$(a) case. Electric field pointing into the conductor strip from the RHM indicates that the charge on the upper part of the strip is negative. However, electric field pointing into the strip from the LHM means the charge on the lower part of the strip is positive. This previously seen behavior of the charge is not inconsistent with a single surface current flow $\mathbf{J}_s$. because, for argument sake, if the bottom charge flows is the $+\hat{z}$ direction, and the top charge in the $-\hat{z}$ direction, they will add and produce a net current. Figs. 5 (a) — (d) show field arrow plots of $\mathbf{H}_t$, the magnetic field in the plane transverse to the z-direction. Again we show the nominal $\bar{\bar{\varepsilon}}$(a) case [Figs. 5(a) and (b) provide, respectively, the actual arrow distribution and the scaled distribution in (161)] and the $\bar{\bar{\varepsilon}}$(d) case [in Figs. 5 (c) and (d)]. (For this frequency at 0.2 GHz, and the frequencies to follow, (161) has been used with $E_{av}$ = 100 V/m and $H_{av}$ = 0.1 amps/m.)

At the second frequency f = 2.0 GHz, we consider the permittivity tensor cases

$$\bar{\bar{\varepsilon}}(a) = \begin{bmatrix} \varepsilon_n & 0 & 0 \\ 0 & \varepsilon_n & 0 \\ 0 & 0 & \varepsilon_n \end{bmatrix} ; \bar{\bar{\varepsilon}}(b) = \begin{bmatrix} \varepsilon_n & 0 & 0 \\ 0 & 4\varepsilon_n & 0 \\ 0 & 0 & \varepsilon_n \end{bmatrix} ; \bar{\bar{\varepsilon}}(c) = \begin{bmatrix} \varepsilon_n & 0 & 0 \\ 0 & \varepsilon_n & 0 \\ 0 & 0 & 4\varepsilon_n \end{bmatrix} ; \bar{\bar{\varepsilon}}(d) = \begin{bmatrix} \varepsilon_n & 0 & 0 \\ 0 & 4\varepsilon_n & 0 \\ 0 & 0 & 4\varepsilon_n \end{bmatrix}$$

$$\bar{\bar{\varepsilon}}(e) = \begin{bmatrix} 4\varepsilon_n & 0 & 0 \\ 0 & 4\varepsilon_n & 0 \\ 0 & 0 & 4\varepsilon_n \end{bmatrix}$$

(163)

Again, the first tensor $\bar{\bar{\varepsilon}}$(a) represents isotropy, and the next three tensors [$\bar{\bar{\varepsilon}}$(b), $\bar{\bar{\varepsilon}}$(c) and $\bar{\bar{\varepsilon}}$(d)] are anisotropic, so their field distribution deviations from $\bar{\bar{\varepsilon}}$(a) show the effect of anisotropy. Finally, the last tensor $\bar{\bar{\varepsilon}}$(e), returns to the isotropic situation. Propagation constant is complex for these five cases, being $\alpha$(a), $\beta$(a) = 1.133, 1.424; $\alpha$(b), $\beta$(b) = 1.999, 2.338; $\alpha$(c), $\beta$(c) = 1.226, 1.692; $\alpha$(d), $\beta$(d) = 1.722, 2.568; $\alpha$(e), $\beta$(e) = 1.728, 2.428. Surface current expansion coefficients (complex, x and z components [see (139)]) for the cases are, respectively, $\{a_{e1}, b_{e1}\}$ = {(- 3.273 × 10$^{-3}$,- 6.645 × 10$^{-3}$); (1,0)}, {(- 1.685 × 10$^{-2}$, - 2.973 × 10$^{-2}$); (1,0)}, {(- 6.278 × 10$^{-3}$, - 9.416 × 10$^{-3}$), (1,0)}, {(- 1.602 × 10$^{-2}$, - 3.243 × 10$^{-2}$), (1,0)}, {(- 7.989 × 10$^{-3}$, - 3.058 × 10$^{-2}$), (1,0)}. Because we are modeling the problem with the LHM lossless, as well as no losses in the strip, bottom ground plane, top cover, or side wall conductors, the presence of a finite $\alpha$ means that some of the power must be flowing in the x-direction as bulk or surface waves as we saw at one tenth of the frequency value before. Confirmation of this behavior comes from examining the electric magnitude E distributions in Figs. 6 (a) — (e) [in color], which show marked intensity above the interface but hugging it along much of the surface for cases $\bar{\bar{\varepsilon}}(a)$, $\bar{\bar{\varepsilon}}(c)$, and $\bar{\bar{\varepsilon}}(e)$ (see further discussion of this subject at the section s end). Figs. 6 (a) and (c) have E similar in shape in that a substantial intensity exists below the strip all the way to the ground plane, just above the strip, and to either side of the strip just above the interface. Figs. 6 (b), (d) and (e) are similar because substantial intensity exists between the strip and ground plane, and above the strip in a half bubble shape. Figs. 7(a) — (e) [in color] show the magnetic field magnitude H. Figs. 7 (a) — (d) show substantial intensity along the interface and above it, again lending strength to the argument that wave propagation normal to the guided wave direction is occurring. All Fig. 7 cases show significant distribution shape variation from one case to another. Figs. 8 (a) — (e) show transverse electric field $\mathbf{E}_t$ arrow plots, in scaled format. Figs. 9 (a) — (e) show transverse magnetic field $\mathbf{H}_t$ arrow plots, in scaled format.

One of the more easily identifiable trends, amongst the anisotropic effects seen above, caused by the change from isotropy to anisotropy, is the enhancement of the E field distribution in the y direction for those tensors which have enhanced the corresponding tensor element [ see $\bar{\bar{\varepsilon}}(b)$ and $\bar{\bar{\varepsilon}}(d)$]. When the last tensor $\bar{\bar{\varepsilon}}(e)$ returns the LHM to isotropy, one sees a field pattern similar to that in $\bar{\bar{\varepsilon}}(a)$ (the initial isotropy, but with lower nominal permittivity value than the final tensor), but with the added effect of enhanced E field behavior in the y direction above the strip, as seen in the intervening anisotropic cases.

To gain some idea of the vast differences between LHM and RHM substrates affecting the field distribution, the last figure after the LHM figures, Fig. 17 gives the electric field distributions for a RHM substrate with the same nominal permittivity magnitude value as the LHM, making its $\varepsilon_n = 2.5$ ($\mu = 1.0$ nonmagnetic) at f = 2.0 GHz. Again five tensors are treated as in (163) (here the multiplicative factor used was 2). One notices that in all the cases shown, $\bar{\bar{\varepsilon}}(a)$ —$\bar{\bar{\varepsilon}}(e)$, the primary feature is the preponderance of electric field magnitude in the RHM substrate and below the metal strip. Secondary feature, arising from the presence of the inflated y tensor element $\varepsilon_{yy}$, is the notch effect, that is, the appearance of a dimple or notch between the edges of the strip above the substrate corresponding to a field magnitude reduction, which is also associated with enhancement of the field magnitude below the interface under the strip. Comparison of the cases in Fig. 17 with those in Fig. 6 shows that a LHM substrate allows vast changes in the field distributions, and that its anisotropy directly affects the field distribution.

Finally, at the third frequency f = 20 GHz, we consider the permittivity tensor cases

$$\bar{\bar{\varepsilon}}(a) = \begin{bmatrix} \varepsilon_n & 0 & 0 \\ 0 & \varepsilon_n & 0 \\ 0 & 0 & \varepsilon_n \end{bmatrix}; \bar{\bar{\varepsilon}}(b) = \begin{bmatrix} \varepsilon_n & 0 & 0 \\ 0 & 4\varepsilon_n & 0 \\ 0 & 0 & \varepsilon_n \end{bmatrix}; \bar{\bar{\varepsilon}}(c) = \begin{bmatrix} \varepsilon_n & 0 & 0 \\ 0 & 4\varepsilon_n & 0 \\ 0 & 0 & 4\varepsilon_n \end{bmatrix}; \bar{\bar{\varepsilon}}(d) = \begin{bmatrix} 4\varepsilon_n & 0 & 0 \\ 0 & 4\varepsilon_n & 0 \\ 0 & 0 & \varepsilon_n \end{bmatrix}$$

$$\bar{\bar{\varepsilon}}(e) = \begin{bmatrix} 4\varepsilon_n & 0 & 0 \\ 0 & 4\varepsilon_n & 0 \\ 0 & 0 & 4\varepsilon_n \end{bmatrix}$$

(164)

As in the previous frequency cases, here again one starts out with isotropy, and finishes with isotropy for this frequency as in the last case. Deviation from these two bounding cases demonstrates the effects of anisotropy. Propagation constant reverts back to a pure phase characteristic for all but the second and third cases in these five cases, being β(a) = 2.394; α(b), β(b) = $1.324 \times 10^{-3}$, 3.653; α(c), β(c) = $1.630 \times 10^{-3}$, 3.721; β(d) = 3.699; β(e) = 3.730. Surface current expansion coefficients (complex, x and z components [see (139)]) for the cases are, respectively, $\{a_{e1}, b_{e1}\}$ = {(0, - 0.5338); (1,0)}, {($1.110 \times 10^{-3}$, - 0.4390); (1,0)}, {(- $2.505 \times 10^{-3}$, - 0.2248), (1,0)}, {((1,0), (0, 0.5368)}, {(0, - 0.5371), (1,0)}. All of the electric magnitudes E look different from each other as seen in Figs. 10 (a) — (e) [in color]. But some of the overall trends can be discerned. For example, below the interface intensity maxima and minima occur periodically. This x-variation must be due to the effective wavelength in that direction. Although it must be determined numerically, and its value found by counting two successive maxima and minima in the plot, an estimate can be calculated as $\lambda = \lambda_0/\sqrt{\varepsilon_{xx}}$ = 15 mm/$\sqrt{\varepsilon_{xx}}$ at 20 GHz. For cases (a) and (d), $\sqrt{\varepsilon_{xx}(a)} = \sqrt{\varepsilon_n} = \sqrt{2.5}$ = 1.58 making λ(a) = 9.5 mm, and $\sqrt{\varepsilon_{xx}(d)} = \sqrt{4\varepsilon_n} = 2\sqrt{2.5}$ = 3.16 making λ(d) = 4.75 mm. This translates into about 5 periods for case (a) and 10.5 periods for case (d) in the 2b = 50 mm width. Inspection of Figs. 10 (a) and (d) shows that the actual number is slightly less, roughly 3 1/4 and 7.5. All cases except (b) have the largest intensity around the strip, with significant field below the strip for cases (a) and (d), with case (a) being much more than case (d). Cases (c) — (e) in Fig. 10 have some field just above the strip. Finally, case (b) in Fig. 10 has only a remnant of a field around the strip, but very noticeable field along the interface, extending all the way to both side walls. Case (b) in Fig. 10 appears to be a clear example of a surface wave, although it is not obvious based upon the α/β ratio being so small. However, this could occur if the surface wave is predominantly in the same direction as the guided wave, the z-direction (see section end for more discussion). Figs. 11(a) — (e) [in color] show the magnetic field magnitude H, and like the E field plots the periodicity variation along the x-direction is seen here also, as well as all cases having significant field around the strip. Surface wave characteristic is again clearly demonstrated in Fig. 11 (b). Figs. 12 (a) — (e) show transverse electric field $\mathbf{E}_t$ arrow plots, in scaled format. We notice that in all cases except $\bar{\bar{\varepsilon}}$(b), the arrows immediately above and below the strip point in or out, indicating opposite charges on the two sides of the conductor strip, an observation made earlier in regard to behavior at f = 0.2 GHz. However, for $\bar{\bar{\varepsilon}}$(b), $\mathbf{E}_t$ arrows point upward above and below the strip, indicating positive charge on both sides of the conductor. Figs. 13 (a) — (e) show transverse magnetic field $\mathbf{H}_t$ arrow plots, in scaled format. As seen for the previous frequency case, tensor element enhancement of the y element tends to push the E field distribution to being just below the strip or above the strip. However, at this frequency, the bubble shape is more squished in appearance.

  In order to discern the field variation in the LHM structure in another way, other than the cross-sectional 2D visualization techniques already covered, simple cuts for specific x or y values may be taken. Here an x cut at x = 15 mm will be done, purposely with x › 0 to avoid going through the strip, with the interval broken into 100 points. One case is chosen from each frequency previously studied, to illustrate the cross-sectional field component variations against x. The cases are $\bar{\bar{\varepsilon}}$(d) for f = 0.2 GHz, $\bar{\bar{\varepsilon}}$(c) for f = 2 GHz, and $\bar{\bar{\varepsilon}}$(b) for f = 20 GHz. All results are for $n_x = n_z = 1$, n = 200, except for Fig. 14 (a) and Fig. 15 (a) done with $n_x = n_z = 5$, n = 600 to capture the interfacial $E_x$ continuity better and for Fig. 16 (c) to capture the interfacial $H_x$ continuity better (see Table I for the surface current coefficients used for these exceptions). At f = 0.2 GHz (Fig. 14), other than the sign switches in $E_y$ and $H_y$ which occur because of continuity of normal components of **D** or **B** across the interface at x = 15 mm (magnitude and sign of the difference in values of the normal electric field component on either side of the interface is controlled by $\varepsilon_{top}E_{ytop} = \varepsilon_{LHM}E_{yLHM}$, or the permittivity ratio $1/2\varepsilon_n = -1/5$; magnitude and sign of the difference in values of the normal magnetic field component on either side of the interface is controlled by $\mu_{top}H_{ytop} = \mu_{LHM}H_{yLHM}$, or the permeability ratio $1/\mu_n = -1/2.5$), the field variation is linear (in the LHM substrate) or if nonlinear, having moderate variation. ($E_x$ and $H_x$ [as well as $E_z$ and $H_z$ ]are continuous across the interface due to **E** and **H** tangential component continuity.) At f = 2 GHz (Fig. 15), similar field component behavior to the lower frequency case is seen, with sign switches (and magnitude) of the normal field components obeying **D** or **B** continuity (permittivity ratio $\varepsilon_{top}/\varepsilon_{LHM} = 1/\varepsilon_n = -1/2.5$; again permeability ratio $\mu_{top}/\mu_{LHM} = 1/\mu_n = -1/2.5$). Radically different field behavior occurs at f = 20 GHz (Fig. 16), where exponentially decaying field away from the interface occurs. If we use exponential functional variation as a definition of pure surface wave behavior, then only this third frequency case qualifies as strictly a surface wave. Sign switches and magnitude differences of the normal field components obey permittivity ratio $\varepsilon_{top}/\varepsilon_{LHM} = 1/4\varepsilon_n = -1/10$ and permeability ratio $\mu_{top}/\mu_{LHM} = 1/\mu_n = -1/2.5$. Note that for Figs. 14 — 16, $E$ and $H_i$, i = x, y, the units are $E_y$ (V/m) and $H_y$ (amps/m) which is mks, but because the solution is rf, small signal, it really can be scaled arbitrarily, making the units really arbitrary. The size of the field components for a particular permittivity and permeability tensor pair choice in a computer simulation run are related because they were solved simultaneously, but field sizes may not be directly compared for different tensor cases run separately. Thus Fig. 2 (d), 3 (d) and 14 all correspond to the same $\bar{\bar{\varepsilon}}$ (d) run at f = 0.2 GHz, and may be directly compared in regard to field component magnitudes.

## V. CONCLUSION

In this paper a compete treatment of the theoretical process for modeling anisotropic left-handed materials (LHM) in guided wave structures has been given. With that formulation, a specific structure had been chosen to study, the single microstrip guided wave device loaded with LHM. Three frequency bands were identified to study, and calculations done at three frequency points within those bands. Field magnitude color distribution plots have been produced for electric E and magnetic H fields. Also, arrow plots have been generated. We have identified propagation of guided waves with and without the generation of surface waves. The surface wave is coupled to the guided wave and extracts energy from the guided wave and results in both a real part of the propagation constant and concomitant field distribution showing side ways extraction of energy. Use of LHM substrates produces field distributions considerably different from ordinary media (right-handed materials, RHM), and it has been shown here that anisotropy can have a significant effect on those distributions. Not studied here explicitly, but known for the nominal structure considered in this paper, backward waves occur (propagation along the z-axis) over part of the frequency spectrum (in regions of the 0.3 — 10 GHz and 30 — 40 GHz bands), by examining the dispersion diagram. Such LHM character of the whole device itself may be of great interest for devices in electronic circuits.

Also seen here is that anisotropy introduced through specific tensor elements can have identifiable aspects displayed in the electromagnetic field distributions. For example, transverse plane anisotropy produces different results than longitudinal anisotropy, and examination of the electric field magnitude distribution shows this behavior. All the field distributions produced in this research on anisotropy were done against a backdrop of isotropy for the initial tensors describing the physical properties of the LHM. The ability to redistribute the volumetric fields of a bulk like wave, or convert a volumetric wave into a surface wave, as found here suggests potential LHM device applications. It may be possible to introduce a new class of control components based upon distortion of the LHM permittivity tensor. Such LHM tensor distortion could be produced by stress or strain, or through an analogy to ferroelectric behavior in materials, for example.

Closely tied in with our study here, the negative refractive property of LHMs has also been examined in a related class of anisotropic crystals [20] — [23] in regard to guided wave propagation in microwave structures [22], [23] which produce newly found asymmetric redistributions of fields. Interesting device applications may result, and this demonstrates that the study of anisotropy in negative refractive materials may be just at its beginning. More interesting physics is expected to be discovered, with other applications.

FIGURE CAPTIONS

1. Cross — sectional diagram of the double layered structure, with the strip sandwiched between the upper RLM and the lower LHM substrate, which is generally a principal axis biaxial LHM crystal for this study. Propagation is perpendicular to the cross — section.
2. Color distribution of the E field for a microstrip guided wave structure, with a LHM substrate $h_1 = 5$ mm thick, and a vacuum overlayer $h_2 = 20$ mm thick, with 50 mm side wall separation. Calculation is done at $f = 0.2$ GHz. Tensor cases have scalar permeability $\mu = \mu_n = -2.5$ and biaxial permittivity which are chosen as (a) $\varepsilon_{xx} = \varepsilon_{yy} = \varepsilon_{zz} = \varepsilon_n = -2.5$ ; (b) $\varepsilon_{xx} = \varepsilon_{zz} = \varepsilon_n$ , $\varepsilon_{yy} = 2\varepsilon_n$ ; (c) $\varepsilon_{yy} = \varepsilon_{zz} = 2\varepsilon_n$ , $\varepsilon_{xx} = \varepsilon_n$ ; (d) $\varepsilon_{xx} = \varepsilon_{yy} = 2\varepsilon_n$ , $\varepsilon_{zz} = \varepsilon_n$ .

3. Color distribution of the H field for a microstrip guided wave structure, with a LHM substrate $h_1 = 5$ mm thick, and a vacuum overlayer $h_2 = 20$ mm thick, with 50 mm side wall separation. Calculation is done at $f = 0.2$ GHz. Tensor cases have scalar permeability $\mu = \mu_n = -2.5$ and biaxial permittivity which are chosen as (a) $\varepsilon_{xx} = \varepsilon_{yy} = \varepsilon_{zz} = \varepsilon_n = -2.5$ ; (b) $\varepsilon_{xx} = \varepsilon_{zz} = \varepsilon_n$ , $\varepsilon_{yy} = 2\varepsilon_n$ ; (c) $\varepsilon_{yy} = \varepsilon_{zz} = 2\varepsilon_n$ , $\varepsilon_{xx} = \varepsilon_n$ ; (d) $\varepsilon_{xx} = \varepsilon_{yy} = 2\varepsilon_n$ , $\varepsilon_{zz} = \varepsilon_n$ .

4. Arrow distribution plots at $f = 0.2$ GHz for the transverse electric $\mathbf{E}_t$ vector for the LHM device in Figs. 2 and 3. (a) unscaled plot for nominal case with $\bar{\bar{\varepsilon}} = \varepsilon_n I$ and $\bar{\bar{\mu}} = \mu_n I$; (b) scaled plot for parameters in (a); (c) unscaled plot with $\varepsilon_{xx} = \varepsilon_{yy} = 2\varepsilon_n$ , $\varepsilon_{zz} = \varepsilon_n = -2.5$; (d) scaled plot for parameters in (c).

5. Arrow distribution plots at $f = 0.2$ GHz for the transverse magnetic $\mathbf{H}_t$ vector for the LHM device in Figs. 2 and 3. (a) unscaled plot for nominal case with $\bar{\bar{\varepsilon}} = \varepsilon_n I$ and $\bar{\bar{\mu}} = \mu_n I$; (b) scaled plot for parameters in (a); (c) unscaled plot with $\varepsilon_{xx} = \varepsilon_{yy} = 2\varepsilon_n$ , $\varepsilon_{zz} = \varepsilon_n = -2.5$; (d) scaled plot for parameters in (c).

6. Color distribution of the E field for a microstrip guided wave structure, with a LHM substrate $h_1 = 5$ mm thick, and a vacuum overlayer $h_2 = 20$ mm thick, with 50 mm side wall separation. Calculation is done at $f = 2.0$ GHz. Tensor cases have scalar permeability $\mu = \mu_n = -2.5$ and biaxial permittivity which are chosen as (a) $\varepsilon_{xx} = \varepsilon_{yy} = \varepsilon_{zz} = \varepsilon_n = -2.5$ ; (b) $\varepsilon_{xx} = \varepsilon_{zz} = \varepsilon_n$ , $\varepsilon_{yy} = 4\varepsilon_n$ ; (c) $\varepsilon_{xx} = \varepsilon_{yy} = \varepsilon_n$ , $\varepsilon_{zz} = 4\varepsilon_n$ ; (d) $\varepsilon_{yy} = \varepsilon_{zz} = 4\varepsilon_n$ , $\varepsilon_{xx} = \varepsilon_n$ ; (e) $\varepsilon_{xx} = \varepsilon_{yy} = \varepsilon_{zz} = 4\varepsilon_n$.

7. Color distribution of the H field for a microstrip guided wave structure, with a LHM substrate $h_1 = 5$ mm thick, and a vacuum overlayer $h_2 = 20$ mm thick, with 50 mm side wall separation. Calculation is done at $f = 2.0$ GHz. Tensor cases have scalar permeability $\mu = \mu_n = -2.5$ and biaxial permittivity which are chosen as (a) $\varepsilon_{xx} = \varepsilon_{yy} = \varepsilon_{zz} = \varepsilon_n = -2.5$ ; (b) $\varepsilon_{xx} = \varepsilon_{zz} = \varepsilon_n$ , $\varepsilon_{yy} = 4\varepsilon_n$ ; (c) $\varepsilon_{xx} = \varepsilon_{yy} = \varepsilon_n$ , $\varepsilon_{zz} = 4\varepsilon_n$ ; (d) $\varepsilon_{yy} = \varepsilon_{zz} = 4\varepsilon_n$ , $\varepsilon_{xx} = \varepsilon_n$ ; (e) $\varepsilon_{xx} = \varepsilon_{yy} = \varepsilon_{zz} = 4\varepsilon_n$.

8. Arrow distribution plots at $f = 2.0$ GHz for the transverse electric $\mathbf{E}_t$ vector for the LHM device in Figs. 2 and 3. All plots are scaled - tensor cases have scalar permeability $\mu = \mu_n = -2.5$ and biaxial permittivity which are chosen as (a) $\varepsilon_{xx} = \varepsilon_{yy} = \varepsilon_{zz} = \varepsilon_n = -2.5$ ; (b) $\varepsilon_{xx} = \varepsilon_{zz} = \varepsilon_n$ , $\varepsilon_{yy} = 4\varepsilon_n$ ; (c) $\varepsilon_{xx} = \varepsilon_{yy} = \varepsilon_n$ , $\varepsilon_{zz} = 4\varepsilon_n$ ; (d) $\varepsilon_{yy} = \varepsilon_{zz} = 4\varepsilon_n$ , $\varepsilon_{xx} = \varepsilon_n$ ; (e) $\varepsilon_{xx} = \varepsilon_{yy} = \varepsilon_{zz} = 4\varepsilon_n$.

9. Arrow distribution plots at $f = 2.0$ GHz for the transverse magnetic $\mathbf{H}_t$ vector for the LHM device in Figs. 2 and 3. All plots are scaled - tensor cases have scalar permeability $\mu = \mu_n = -2.5$ and biaxial permittivity which are chosen as (a) $\varepsilon_{xx} = \varepsilon_{yy} = \varepsilon_{zz} = \varepsilon_n = -2.5$ ; (b) $\varepsilon_{xx} = \varepsilon_{zz} = \varepsilon_n$ , $\varepsilon_{yy} = 4\varepsilon_n$ ; (c) $\varepsilon_{xx} = \varepsilon_{yy} = \varepsilon_n$ , $\varepsilon_{zz} = 4\varepsilon_n$ ; (d) $\varepsilon_{yy} = \varepsilon_{zz} = 4\varepsilon_n$ , $\varepsilon_{xx} = \varepsilon_n$ ; (e) $\varepsilon_{xx} = \varepsilon_{yy} = \varepsilon_{zz} = 4\varepsilon_n$.

10. Color distribution of the E field for a microstrip guided wave structure, with a LHM substrate $h_1 = 5$ mm thick, and a vacuum overlayer $h_2 = 20$ mm thick, with 50 mm side wall separation. Calculation is done at $f = 20$ GHz. Tensor cases have scalar permeability $\mu = \mu_n = -2.5$ and biaxial permittivity which are chosen as (a) $\varepsilon_{xx} = \varepsilon_{yy} = \varepsilon_{zz} = \varepsilon_n = -2.5$ ; (b) $\varepsilon_{xx} = \varepsilon_{zz} = \varepsilon_n$ , $\varepsilon_{yy} = 4\varepsilon_n$ ; (c) $\varepsilon_{yy} = \varepsilon_{zz} = 4\varepsilon_n$ , $\varepsilon_{xx} = \varepsilon_n$ ; (d) $\varepsilon_{xx} = \varepsilon_{yy} = 4\varepsilon_n$ , $\varepsilon_{zz} = \varepsilon_n$ ; (e) $\varepsilon_{xx} = \varepsilon_{yy} = \varepsilon_{zz} = 4\varepsilon_n$.

11. Color distribution of the H field for a microstrip guided wave structure, with a LHM substrate $h_1 = 5$ mm thick, and a vacuum overlayer $h_2 = 20$ mm thick, with 50 mm side wall separation. Calculation is done at $f = 20$ GHz. Tensor cases have scalar permeability $\mu =$ $\mu_n = -2.5$ and biaxial permittivity which are chosen as (a) $\varepsilon_{xx} = \varepsilon_{yy} = \varepsilon_{zz} = \varepsilon_n = -2.5$ ; (b) $\varepsilon_{xx} = \varepsilon_{zz} = \varepsilon_n$ , $\varepsilon_{yy} = 4\varepsilon_n$ ; (c) $\varepsilon_{yy} = \varepsilon_{zz} = 4\varepsilon_n$ , $\varepsilon_{xx} = \varepsilon_n$ ; (d) $\varepsilon_{xx} = \varepsilon_{yy} = 4\varepsilon_n$ , $\varepsilon_{zz} = \varepsilon_n$ ; (e) $\varepsilon_{xx} = \varepsilon_{yy} = \varepsilon_{zz} = 4\varepsilon_n$.

12. Arrow distribution plots at f = 20 GHz for the transverse electric $\mathbf{E}_t$ vector for the LHM device in Figs. 2 and 3. All plots are scaled - tensor cases have scalar permeability $\mu = \mu_n = -2.5$ and biaxial permittivity which are chosen as (a) $\varepsilon_{xx} = \varepsilon_{yy} = \varepsilon_{zz} = \varepsilon_n = -2.5$ ; (b) $\varepsilon_{xx} = \varepsilon_{zz} = \varepsilon_n$ , $\varepsilon_{yy} = 4\varepsilon_n$ ; (c) $\varepsilon_{yy} = \varepsilon_{zz} = 4\varepsilon_n$ , $\varepsilon_{xx} = \varepsilon_n$ ; (d) $\varepsilon_{xx} = \varepsilon_{yy} = 4\varepsilon_n$ , $\varepsilon_{zz} = \varepsilon_n$ ; (e) $\varepsilon_{xx} = \varepsilon_{yy} = \varepsilon_{zz} = 4\varepsilon_n$.

13. Arrow distribution plots at f = 20 GHz for the transverse magnetic $\mathbf{H}_t$ vector for the LHM device in Figs. 2 and 3. All plots are scaled - tensor cases have scalar permeability $\mu = \mu_n = -2.5$ and biaxial permittivity which are chosen as (a) $\varepsilon_{xx} = \varepsilon_{yy} = \varepsilon_{zz} = \varepsilon_n = -2.5$ ; (b) $\varepsilon_{xx} = \varepsilon_{zz} = \varepsilon_n$ , $\varepsilon_{yy} = 4\varepsilon_n$ ; (c) $\varepsilon_{yy} = \varepsilon_{zz} = 4\varepsilon_n$ , $\varepsilon_{xx} = \varepsilon_n$ ; (d) $\varepsilon_{xx} = \varepsilon_{yy} = 4\varepsilon_n$ , $\varepsilon_{zz} = \varepsilon_n$ ; (e) $\varepsilon_{xx} = \varepsilon_{yy} = \varepsilon_{zz} = 4\varepsilon_n$.

14. Field component variations versus y (mm) along the cut x = 15 mm for a microstrip guided wave structure, with a LHM substrate $h_1 = 5$ mm thick, and a vacuum overlayer $h_2 = 20$ mm thick, with 50 mm side wall separation. The frequency is f = 0.2 GHz for permittivity case $\bar{\bar{\varepsilon}}$ (d) with $\varepsilon_{xx} = \varepsilon_{yy} = 2\varepsilon_n$ , $\varepsilon_{zz} = \varepsilon_n = -2.5$. (a) $E_x$ component (b) $E_y$ component (c) $H_x$ component (d) $H_y$ component.

15. Field component variations versus y (mm) along the cut x = 15 mm for the LHM microstrip guided wave structure of Fig.14. The frequency is f = 2 GHz for permittivity case $\bar{\bar{\varepsilon}}$ (c) with $\varepsilon_{xx} = \varepsilon_{yy} = \varepsilon_n = -2.5$, $\varepsilon_{zz} = 4\varepsilon_n$ . (a) $E_x$ component, (b) $E_y$ component, (c) $H_x$ component, (d) $H_y$ component.

16. Field component variations versus y (mm) along the cut x = 15 mm for the LHM microstrip guided wave structure of Fig.14. The frequency is f = 20 GHz for permittivity case $\bar{\bar{\varepsilon}}$ (b) with $\varepsilon_{xx} = \varepsilon_{zz} = \varepsilon_n = -2.5$, $\varepsilon_{yy} = 4\varepsilon_n$. (a) $E_x$ component, (b) $E_y$ component, (c) $H_x$ component, (d) $H_y$ component.

17. Color distribution of the E field for a microstrip guided wave structure, with a RHM dielectric substrate $h_1 = 5$ mm thick, and a vacuum overlayer $h_2 = 20$ mm thick, with 50 mm side wall separation. Calculation is done at f = 2.0 GHz. Tensor cases have scalar permeability $\mu = \mu_n = 1.0$ and biaxial permittivity which are chosen as (a) $\varepsilon_{xx} = \varepsilon_{yy} = \varepsilon_{zz} = \varepsilon_n = 2.5$ ; (b) $\varepsilon_{xx} = \varepsilon_{zz} = \varepsilon_n$ , $\varepsilon_{yy} = 2\varepsilon_n$ ; (c) $\varepsilon_{xx} = \varepsilon_{yy} = \varepsilon_n$ , $\varepsilon_{zz} = 2\varepsilon_n$ ; (d) $\varepsilon_{yy} = \varepsilon_{zz} = 2\varepsilon_n$ , $\varepsilon_{xx} = \varepsilon_n$ ; (e) $\varepsilon_{xx} = \varepsilon_{yy} = \varepsilon_{zz} = 2\varepsilon_n$. $\beta$s are respectively, 1.4052, 1.8188, 1.4053, 1.8198, 1.9219.

TABLE I

Surface Current Coefficients for $n_x = n_z = 5$ and $n = 600$ [see (137) or (139)]
Corresponding to Figs. 13 (a), 14 (a) and 15 (c)

| i | f = 0.2 GHz $\bar{\bar{\varepsilon}}$ (d) $J_{xe}$ $a_{ei}$ | f = 2 GHz $\bar{\bar{\varepsilon}}$ (c) $J_{xe}$ $a_{ei}$ | f = 20 GHz $\bar{\bar{\varepsilon}}$ (b) $J_{xe}$ $a_{ei}$ |
|---|---|---|---|
| 1 | $(1.8351 \times 10^{-3}, 4.3927 \times 10^{-3})$ | $(-1.4428 \times 10^{-3}, -1.0938 \times 10^{-2})$ | $(-6.0329 \times 10^{-3}, -9.5096 \times 10^{-1})$ |
| 2 | $(-2.3666 \times 10^{-4}, -5.5963 \times 10^{-4})$ | $(-1.9032 \times 10^{-4}, 1.3574 \times 10^{-3})$ | $(1.5952 \times 10^{-3}, -2.9726 \times 10^{-1})$ |
| 3 | $(8.4232 \times 10^{-5}, 1.4838 \times 10^{-4})$ | $(5.7020 \times 10^{-5}, -4.3501 \times 10^{-4})$ | $(-9.5447 \times 10^{-3}, 1.2971 \times 10^{-1})$ |
| 4 | $(-1.8509 \times 10^{-5}, -1.4672 \times 10^{-4})$ | $(-3.0738 \times 10^{-5}, 2.6851 \times 10^{-4})$ | $(-2.1657 \times 10^{-2}, 1.3878 \times 10^{-1})$ |
| 5 | $(6.5200 \times 10^{-6}, 4.8388 \times 10^{-5})$ | $(1.7669 \times 10^{-5}, 3.0188 \times 10^{-6})$ | $(2.0023 \times 10^{-4}, 1.7385 \times 10^{-2})$ |

| i | $J_{ze}$ $b_{ei}$ | $J_{ze}$ $b_{ei}$ | $J_{ze}$ $b_{ei}$ |
|---|---|---|---|
| 1 | $(1, 0)$ | $(1, 0)$ | $(-1.7115 \times 10^{-1}, -1.1041 \times 10^{-3})$ |
| 2 | $(1.5983 \times 10^{-1}, -3.5862 \times 10^{-2})$ | $(3.5268 \times 10^{-2}, -6.7151 \times 10^{-2})$ | $(1, 0)$ |
| 3 | $(-3.9399 \times 10^{-2}, 9.9047 \times 10^{-3})$ | $(-7.7496 \times 10^{-3}, 1.7510 \times 10^{-2})$ | $(2.3721 \times 10^{-1}, -1.0963 \times 10^{-2})$ |
| 4 | $(1.3619 \times 10^{-2}, -6.8264 \times 10^{-3})$ | $(5.8341 \times 10^{-4}, -9.0669 \times 10^{-3})$ | $(1.4831 \times 10^{-1}, 3.2868 \times 10^{-2})$ |
| 5 | $(-1.8171 \times 10^{-2}, 7.3621 \times 10^{-5})$ | $(6.0421 \times 10^{-3}, 8.7917 \times 10^{-3})$ | $(5.8148 \times 10^{-1}, 7.2386 \times 10^{-2})$ |

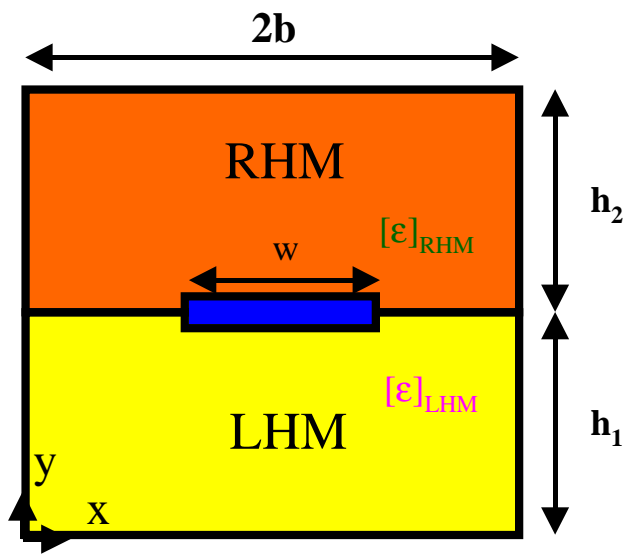

FIGURE 1

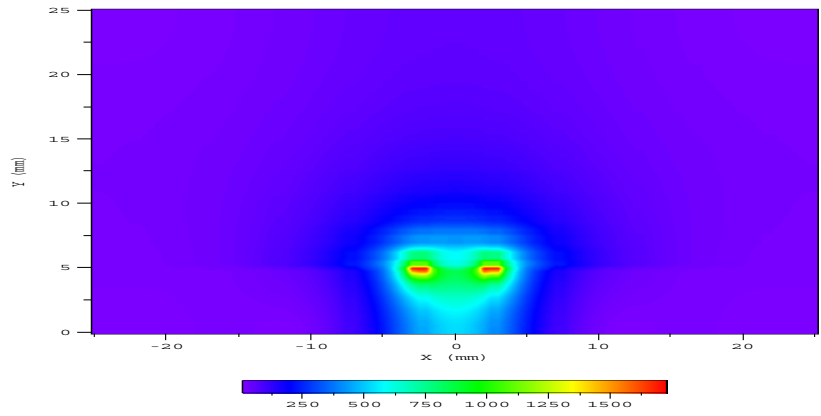

Fig. 2 (a)

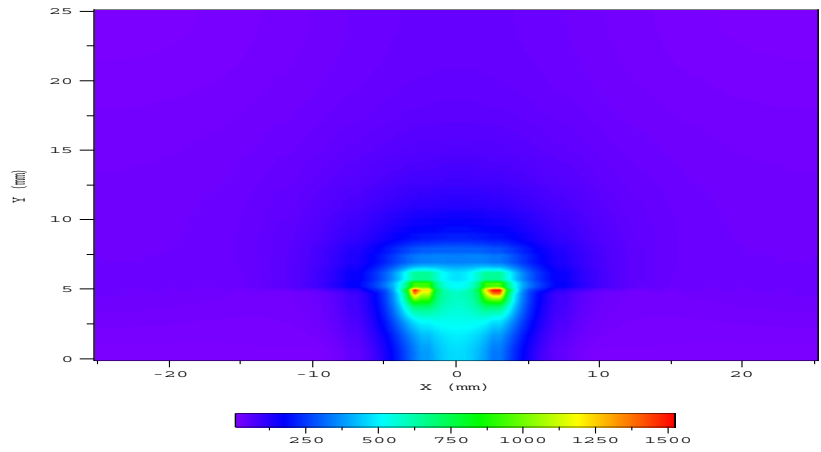

Fig. 2 (b)

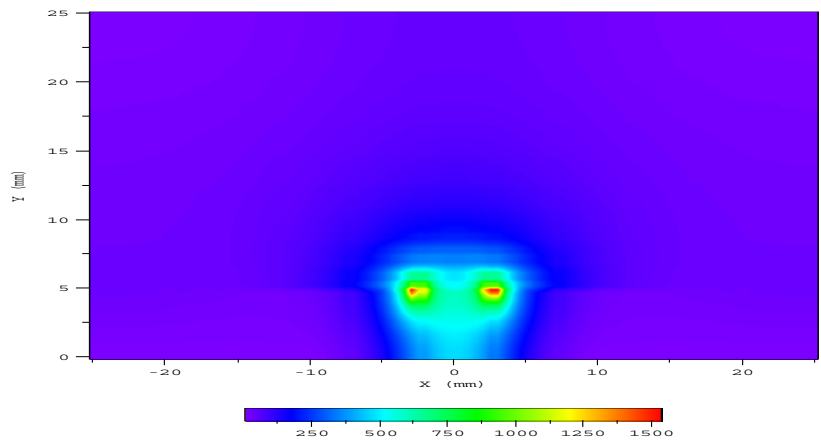

Fig. 2 (c)

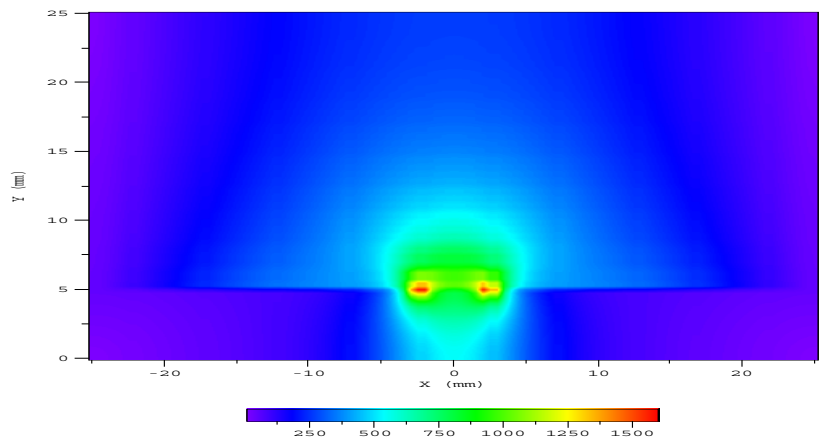

Fig. 2 (d)

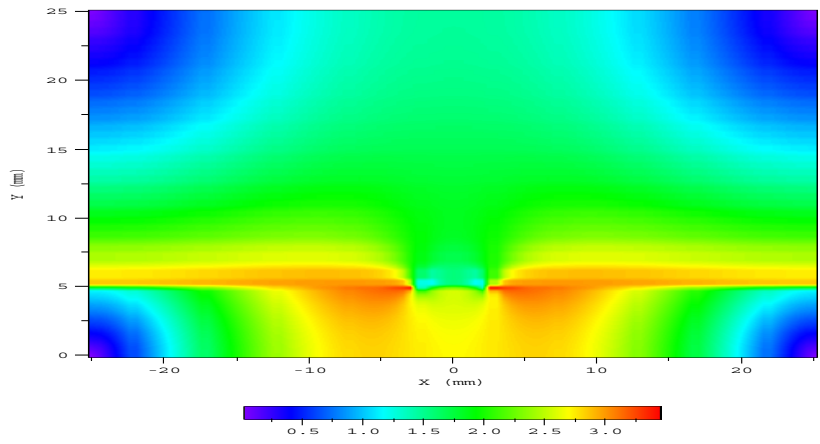
Fig. 3 (a)

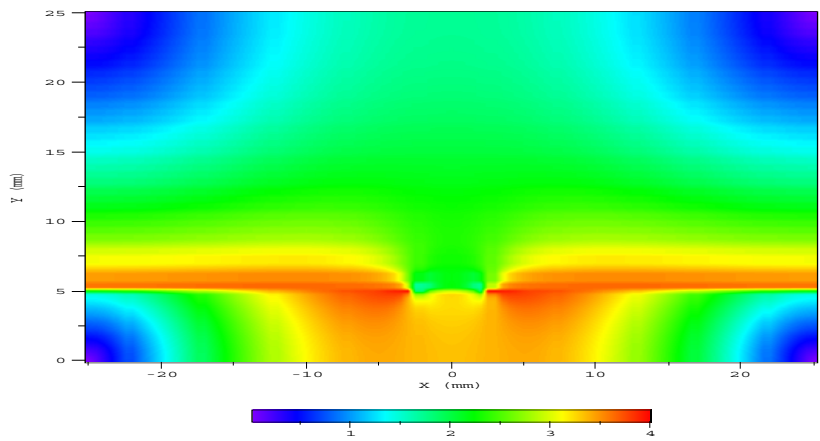
Fig. 3 (b)

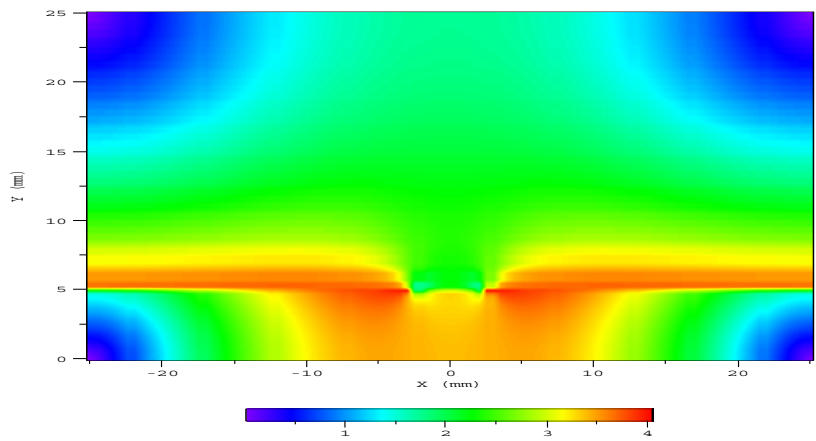
Fig. 3 (c)

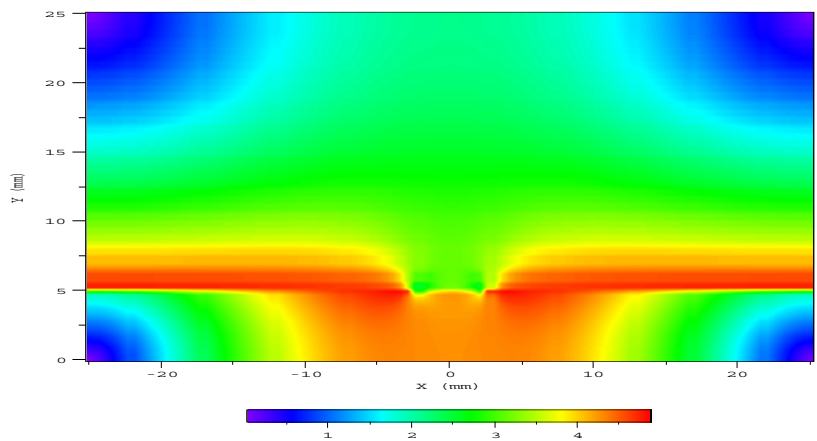
Fig. 3 (d)

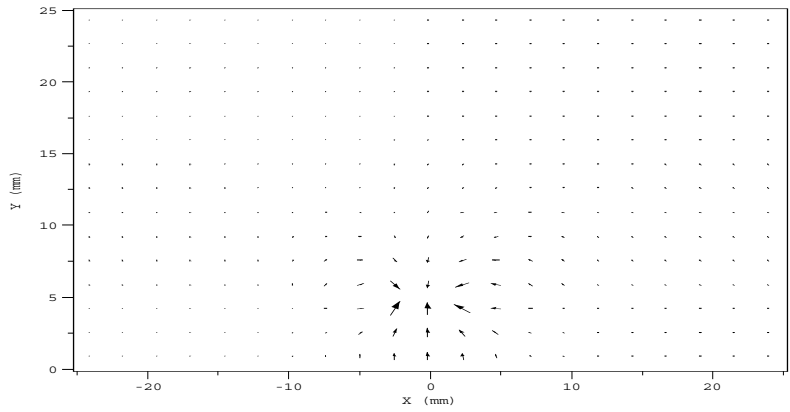

Fig. 4 (a)

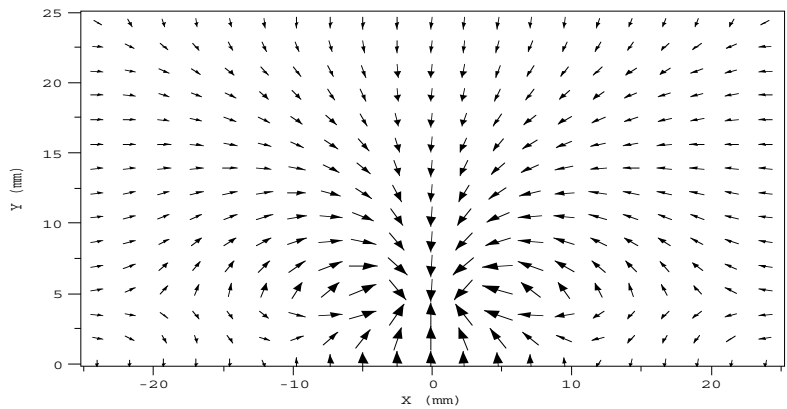

Fig. 4 (b)

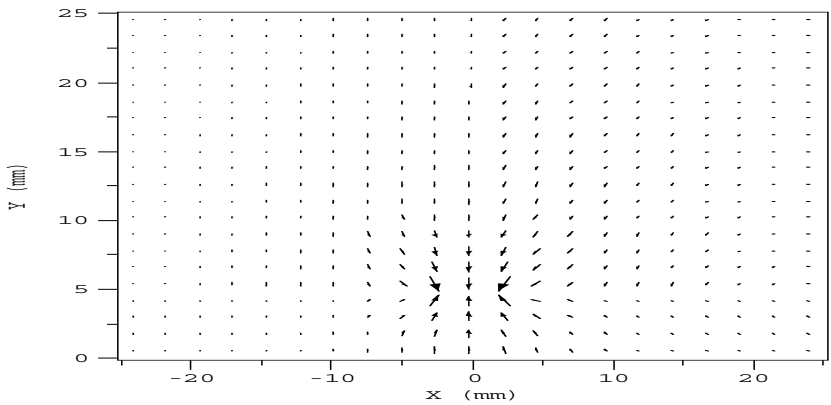

Fig. 4 (c)

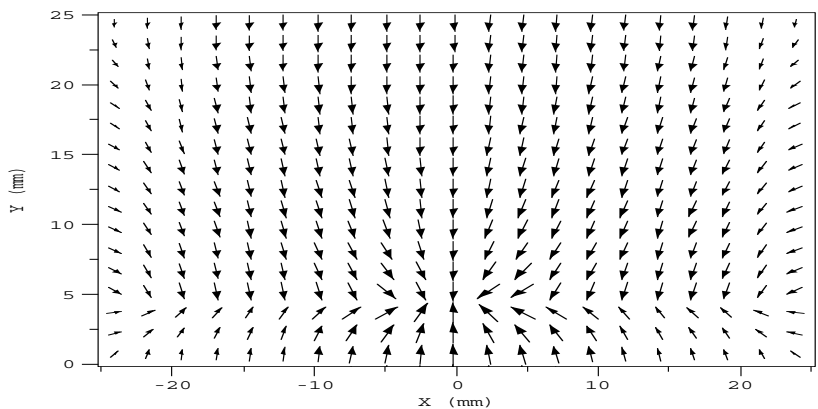

Fig. 4 (d)

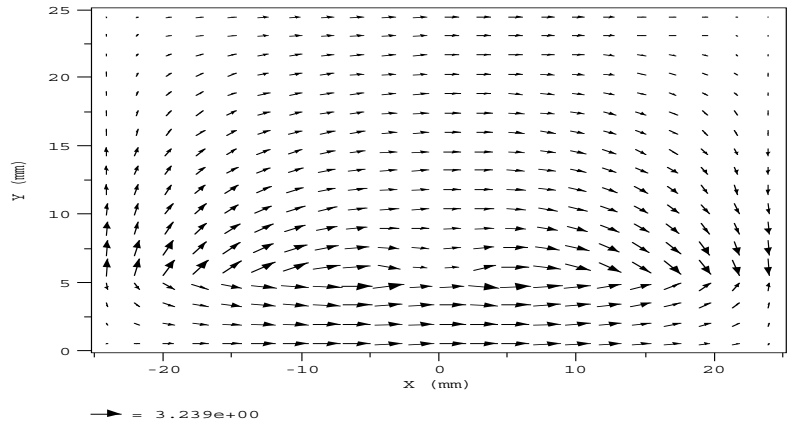

Fig. 5 (a)

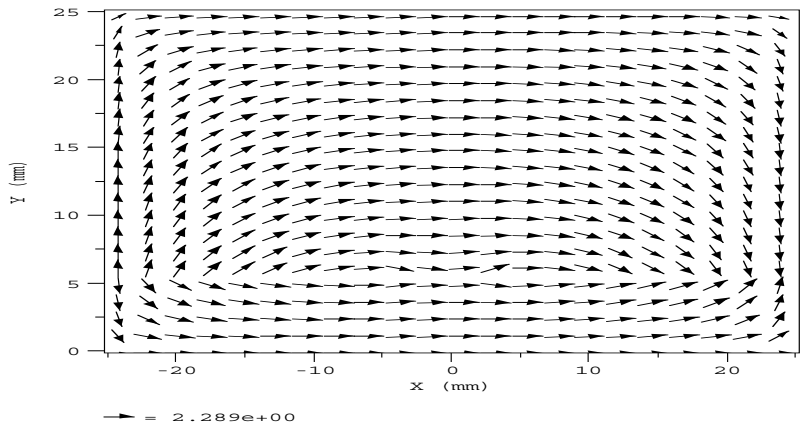

Fig. 5 (b)

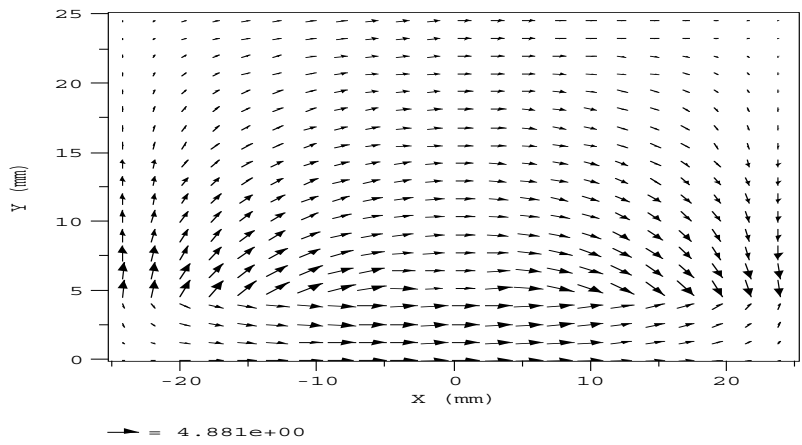

Fig. 5 (c)

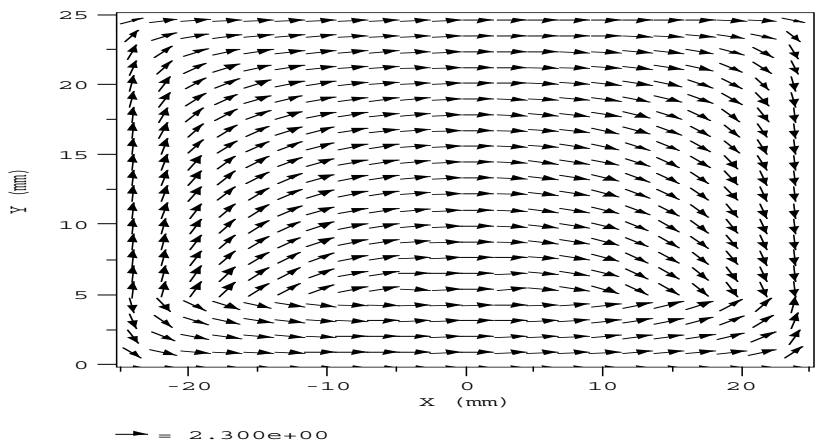

Fig. 5 (d)

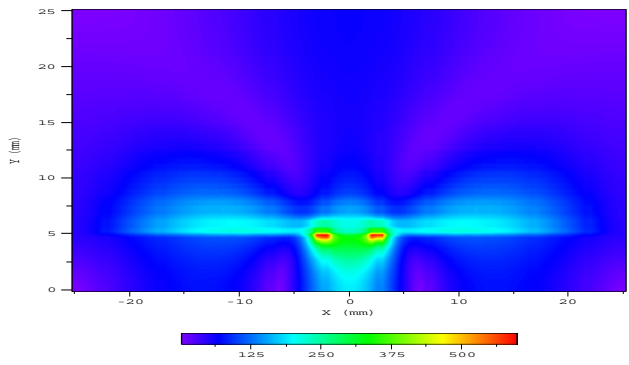

Fig. 6 (a)

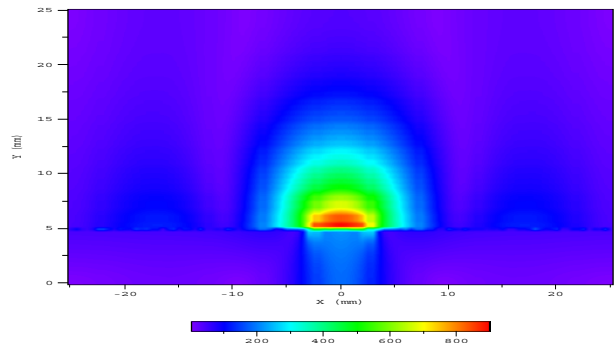

Fig. 6 (b)

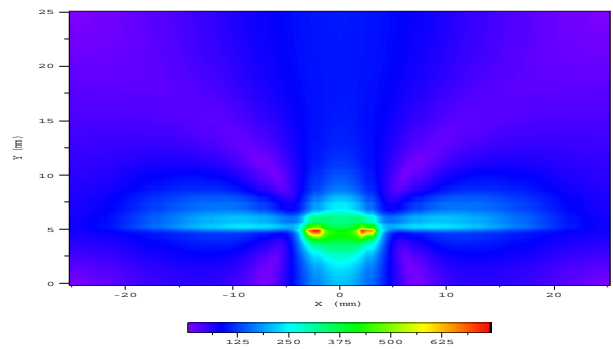

Fig. 6 (c)

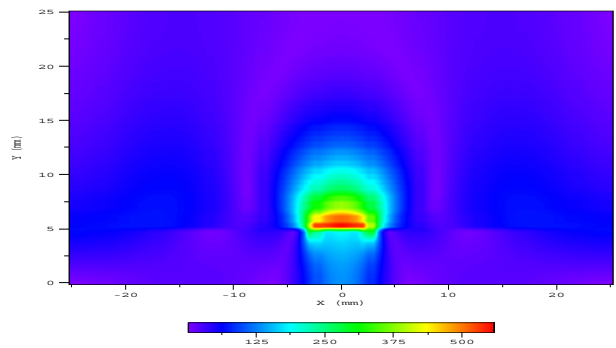

Fig. 6 (d)

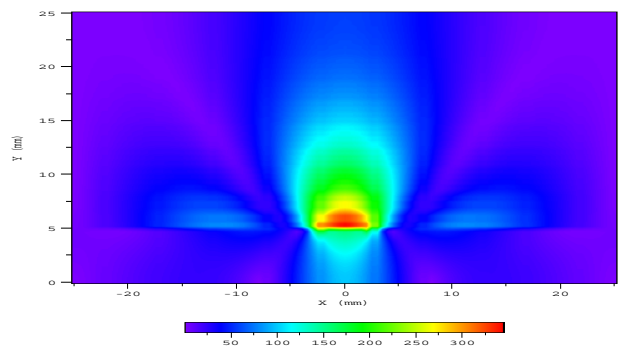

Fig. 6 (e)

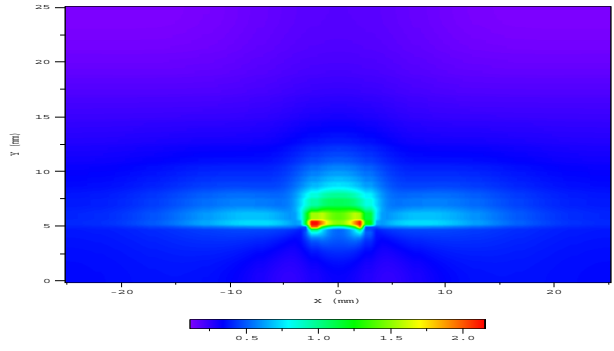

Fig. 7 (a)

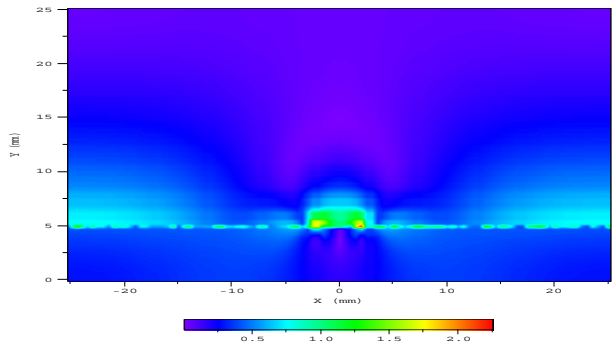

Fig. 7 (b)

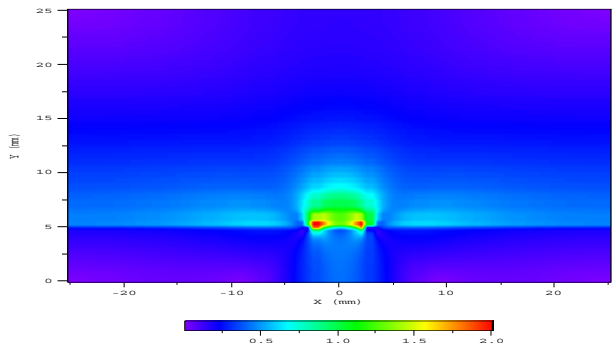

Fig. 7 (c)

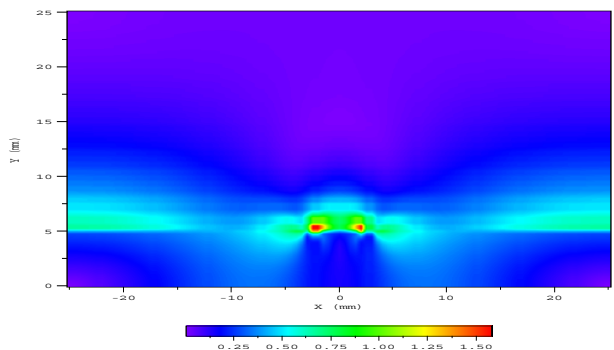

Fig. 7 (d)

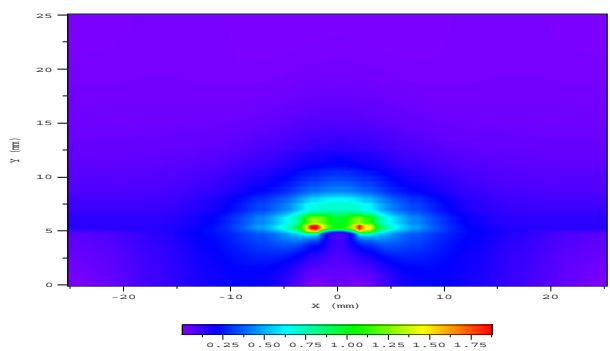

Fig. 7 (e)

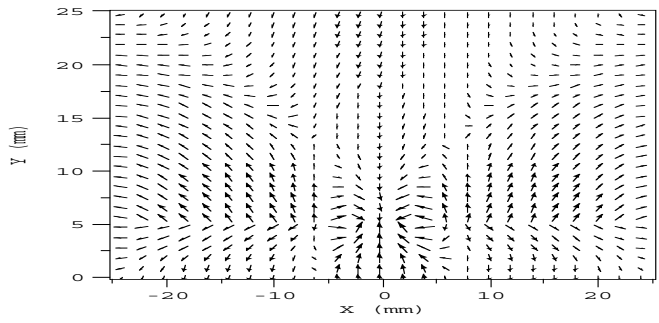

Fig. 8 (a)

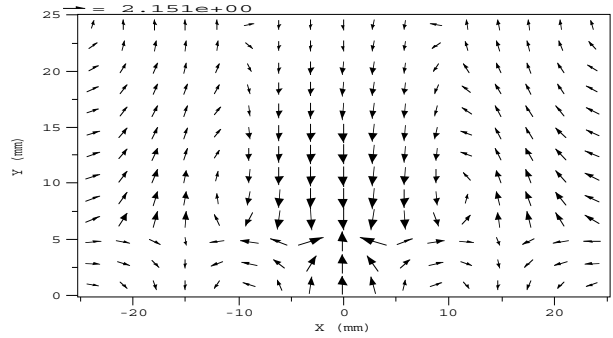

Fig. 8 (b)

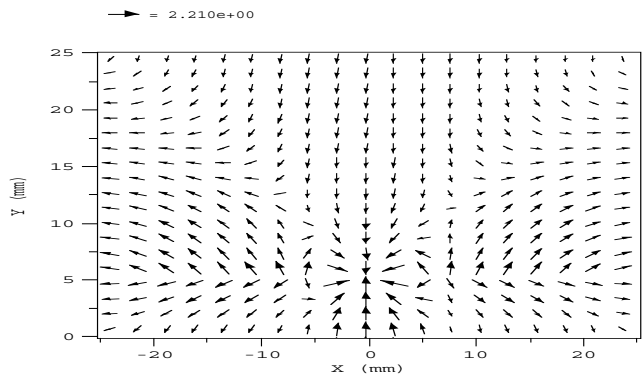

Fig. 8 (c)

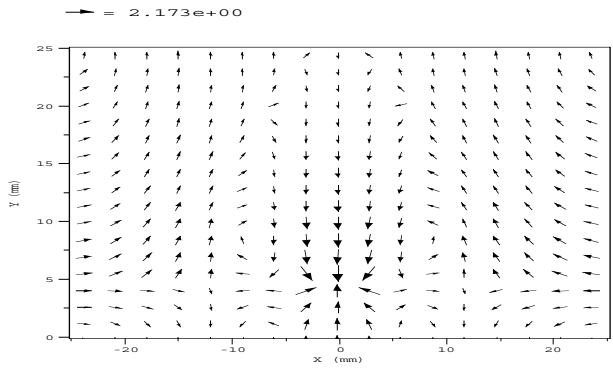

Fig. 8 (d)

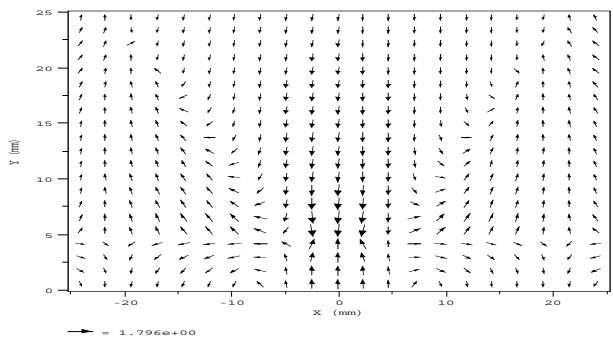

Fig. 8 (e)

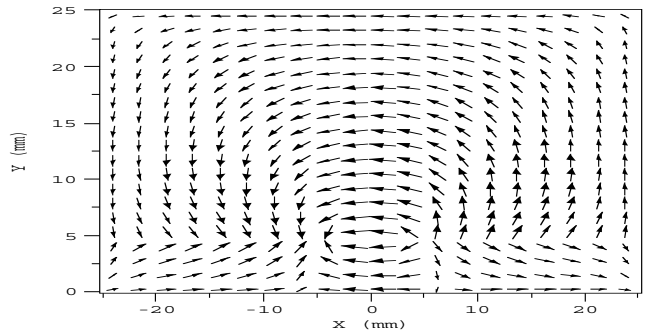

Fig. 9 (a)

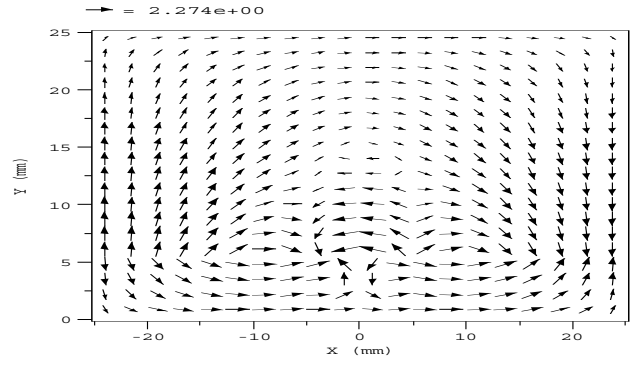

Fig. 9 (b)

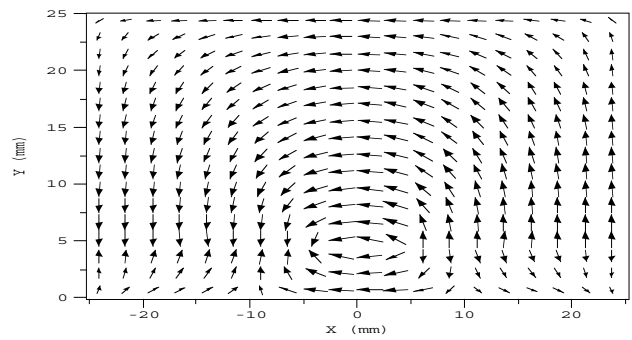

Fig. 9 (c)

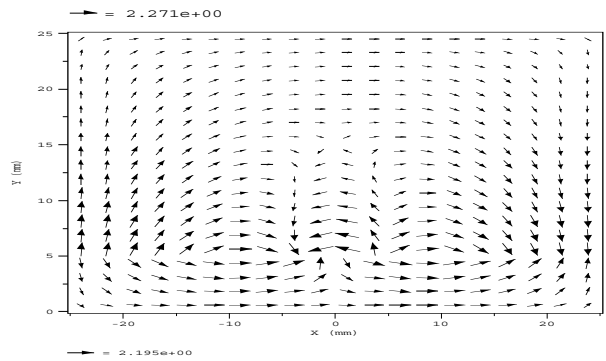

Fig. 9 (d)

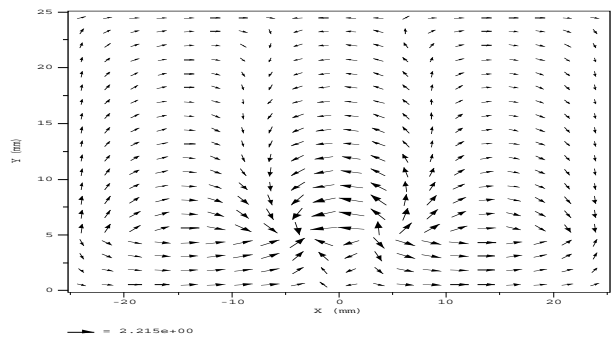

Fig. 9 (e)

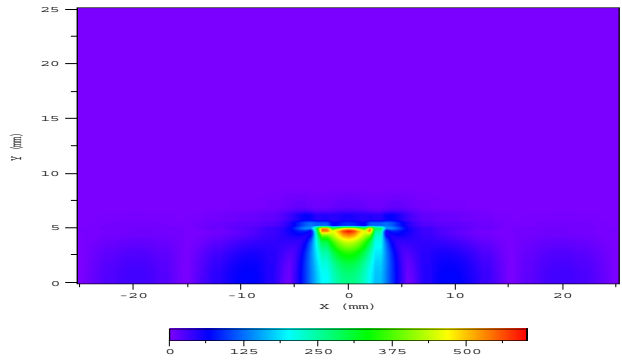

Fig. 10 (a)

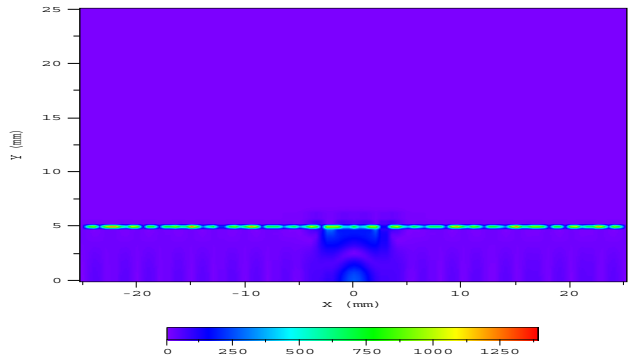

Fig. 10 (b)

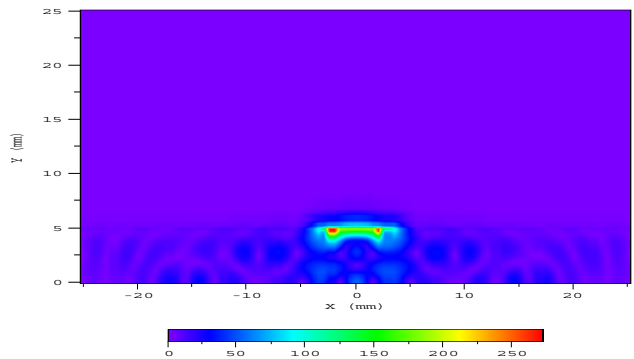

Fig. 10 (c)

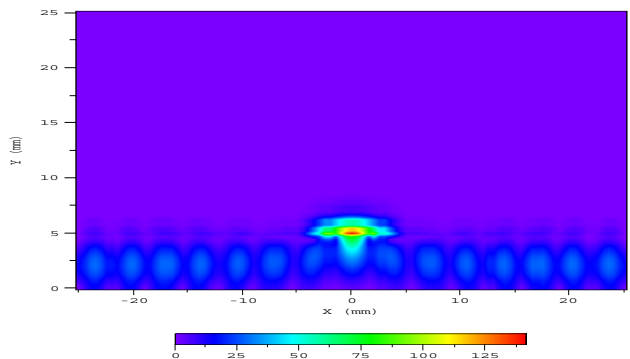

Fig. 10 (d)

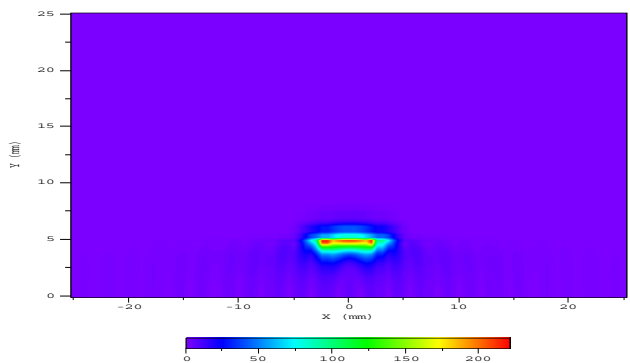

Fig. 10 (e)

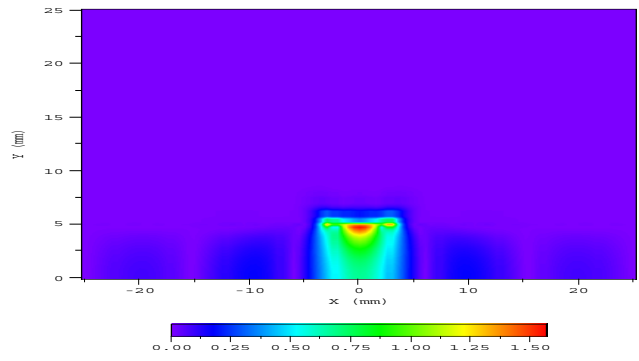

Fig. 11 (a)

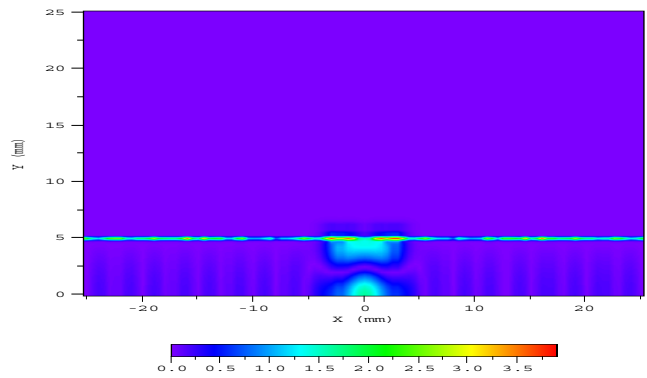

Fig. 11 (b)

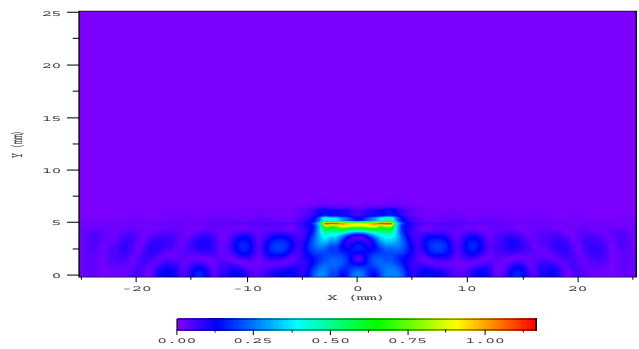

Fig. 11 (c)

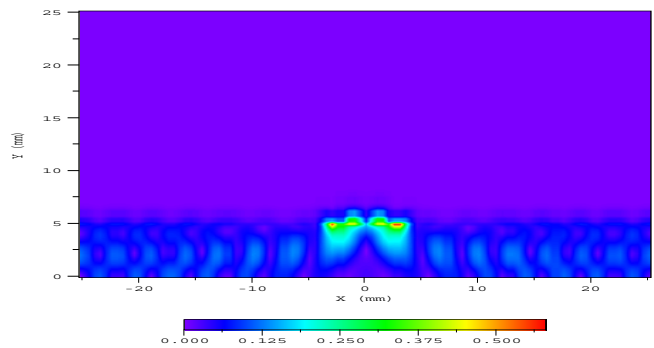

Fig. 11 (d)

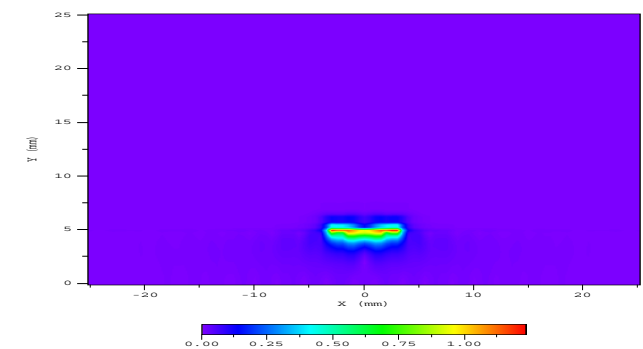

Fig. 11 (e)

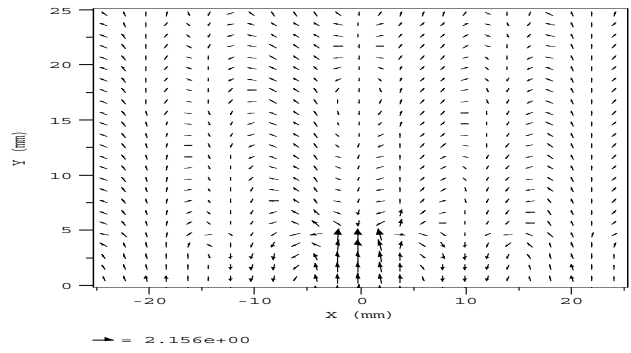

Fig. 12 (a)

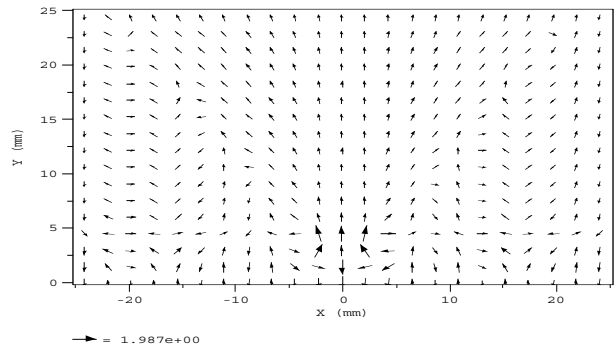

Fig. 12 (b)

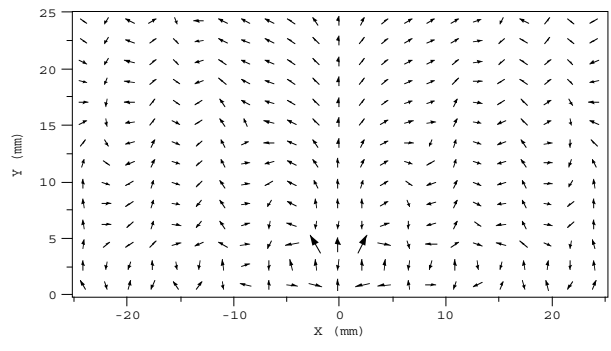

Fig. 12 (c)

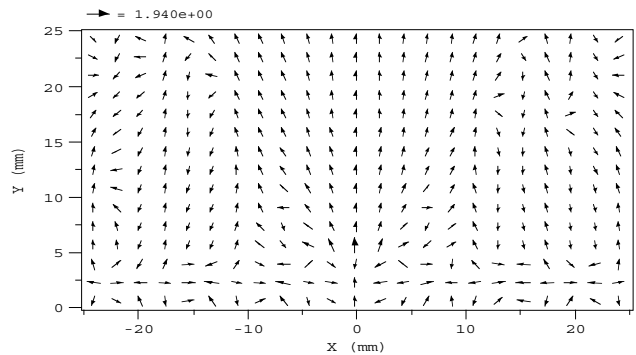

Fig. 12 (d)

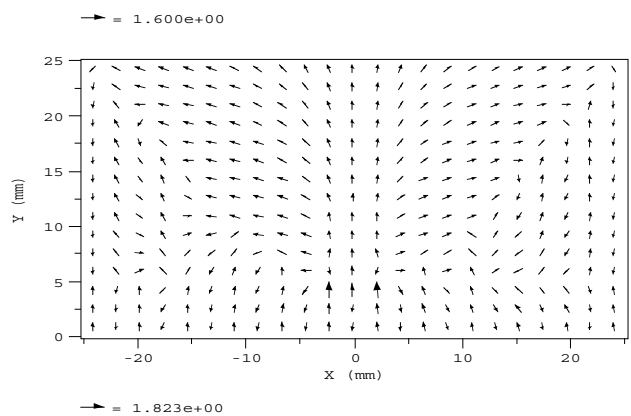

Fig. 12 (e)

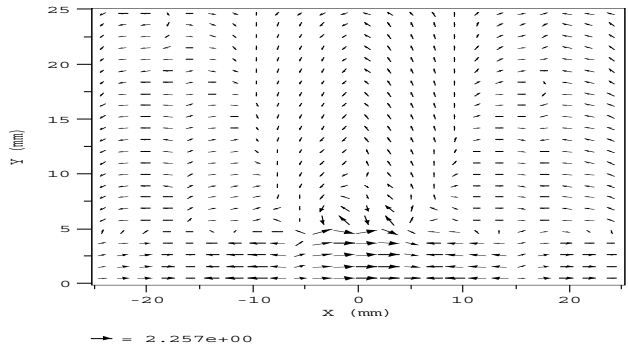

Fig. 13 (a)

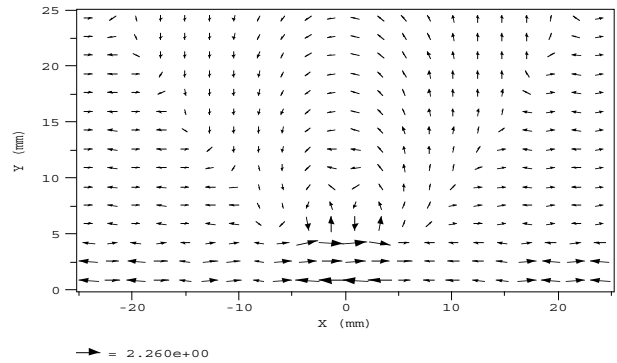

Fig. 13 (b)

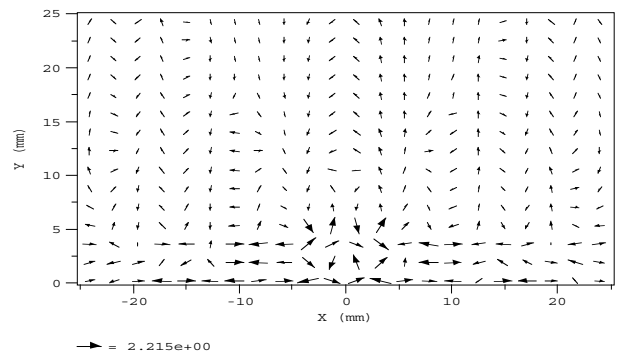

Fig. 13 (c)

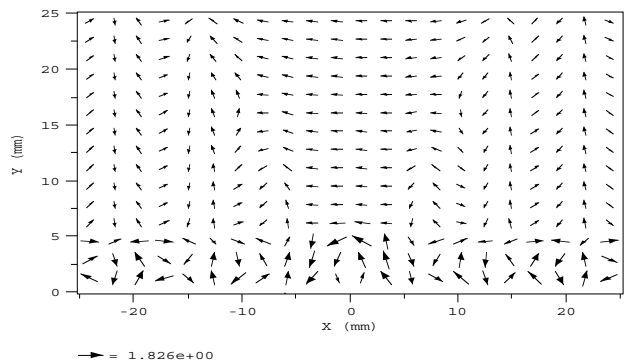

Fig. 13 (d)

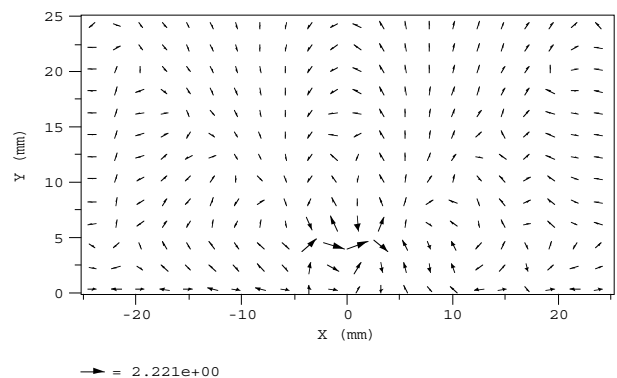

Fig. 13 (e)

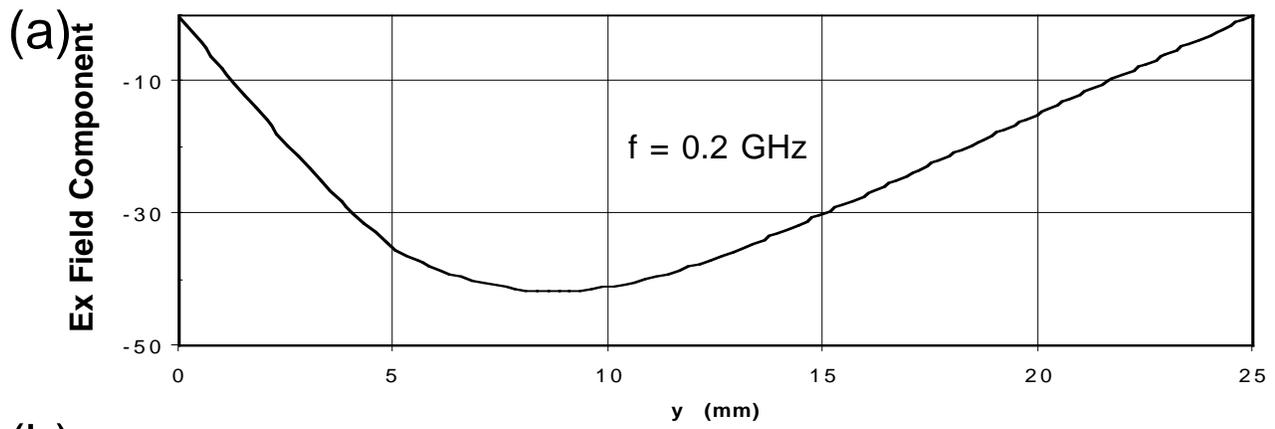

(a)

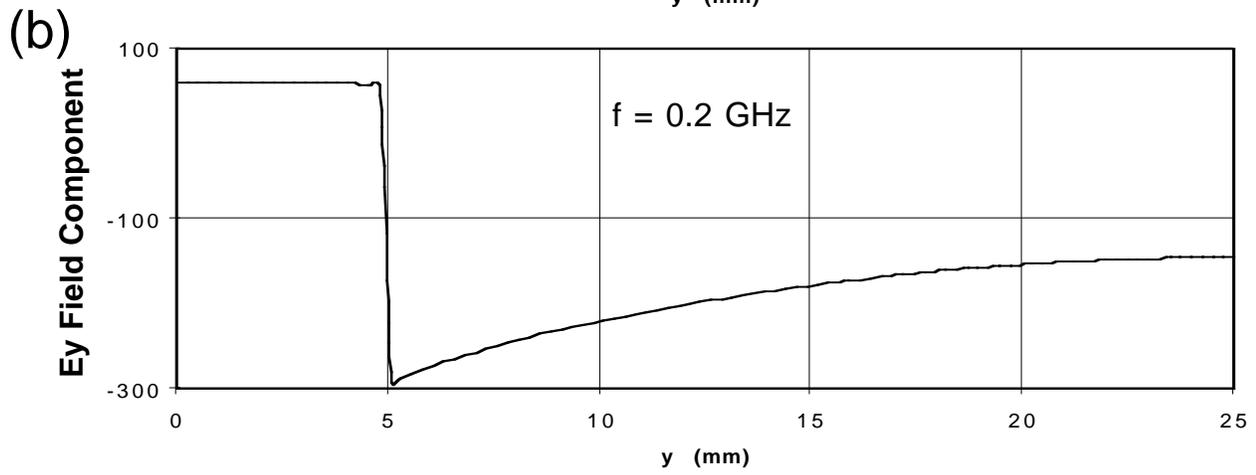

(b)

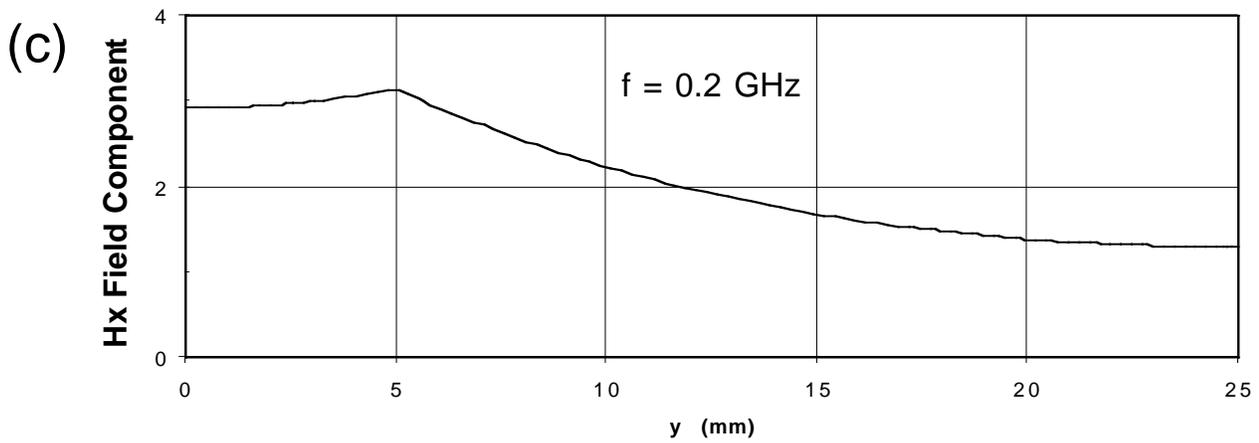

(c)

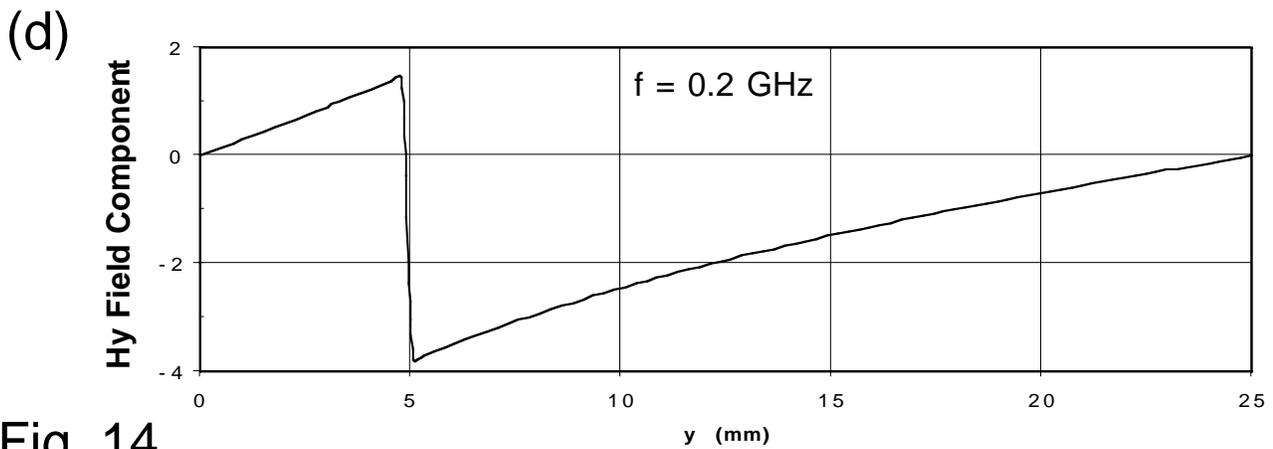

(d)

Fig. 14

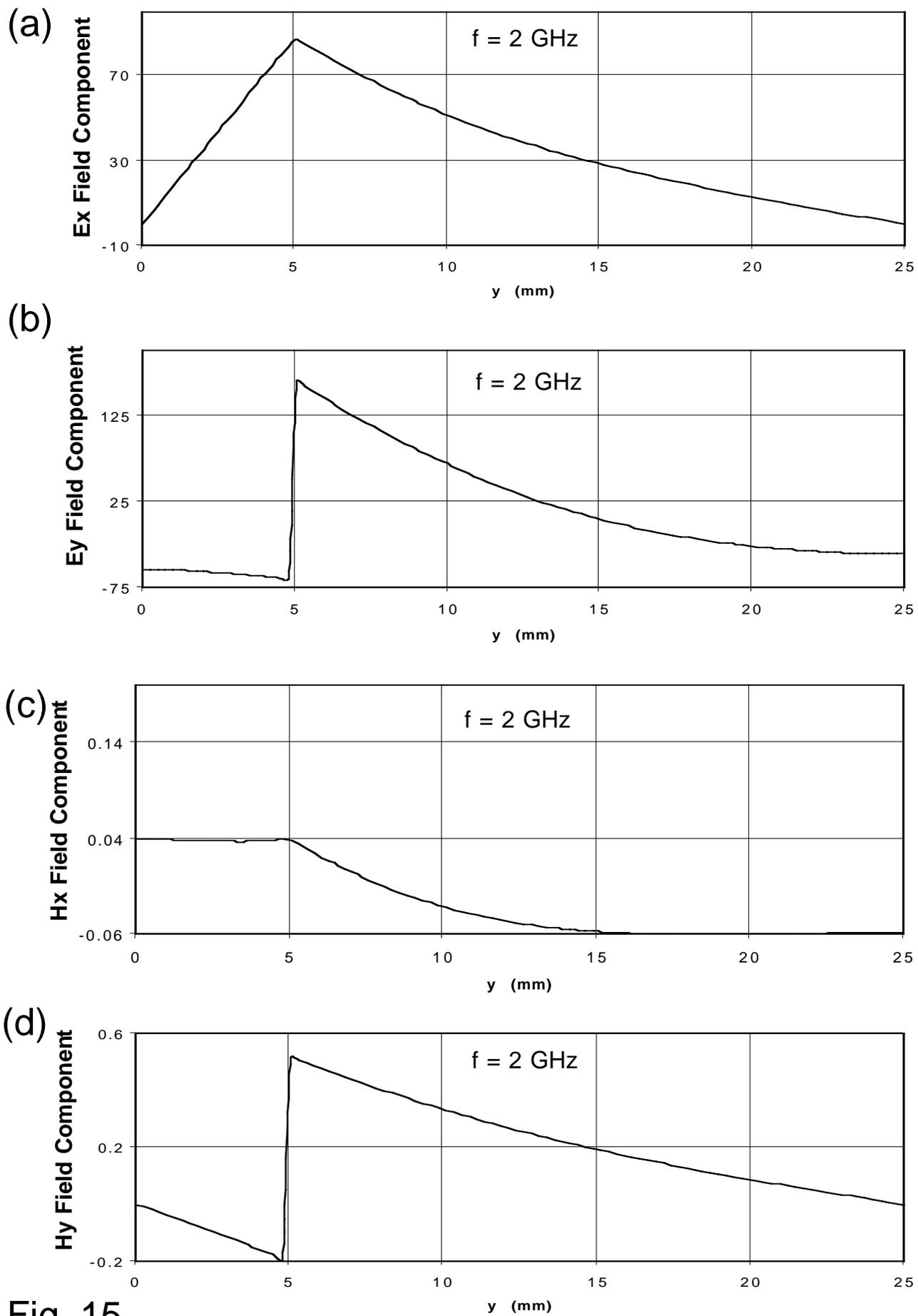

Fig. 15

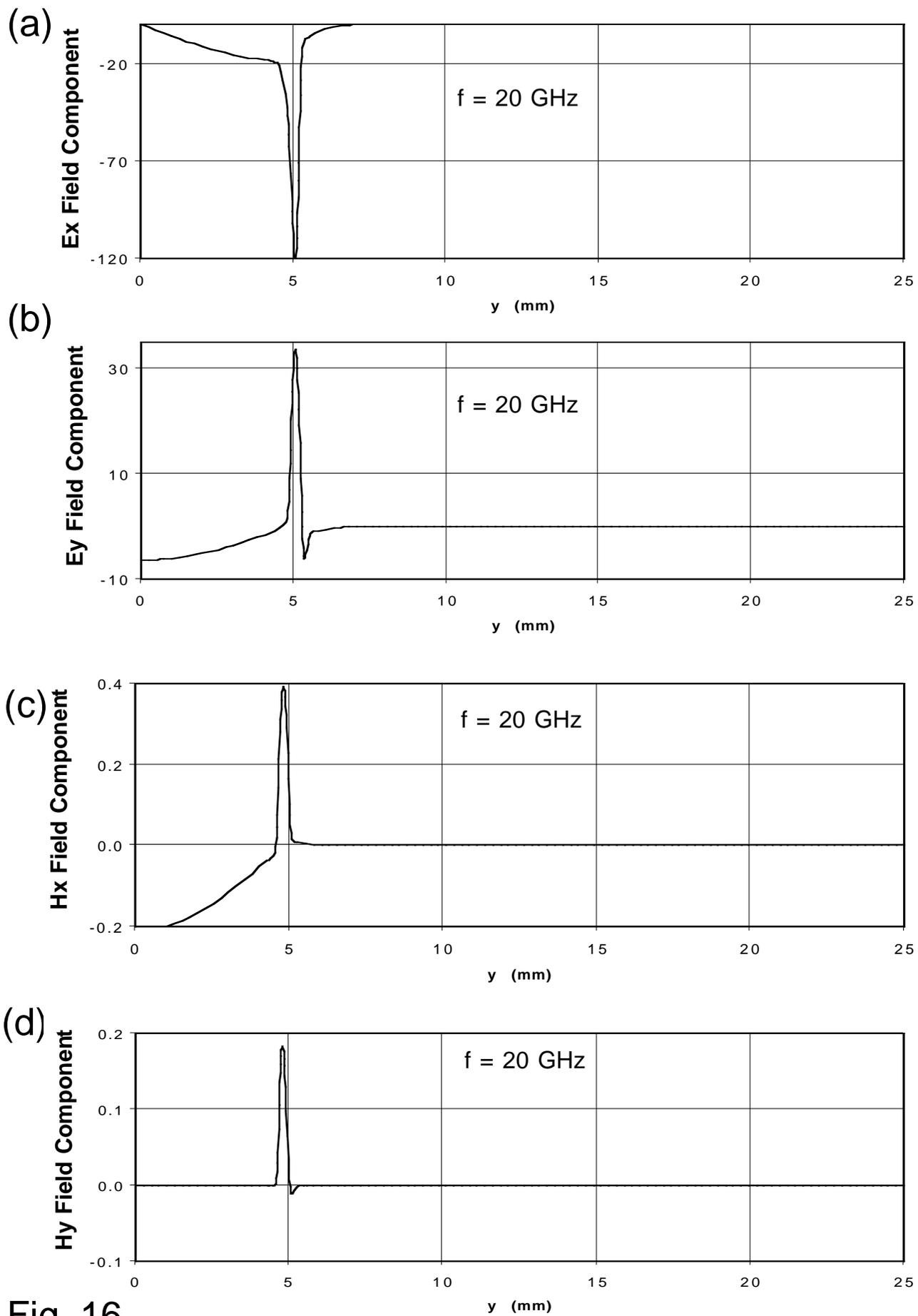

Fig. 16

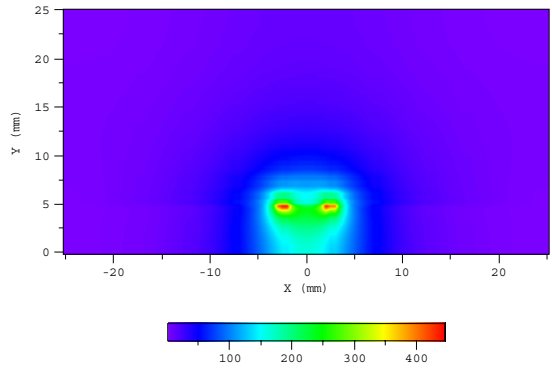

Fig. 17 (a)

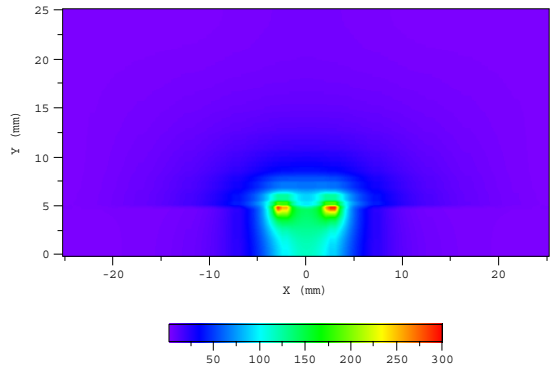

Fig. 17 (b)

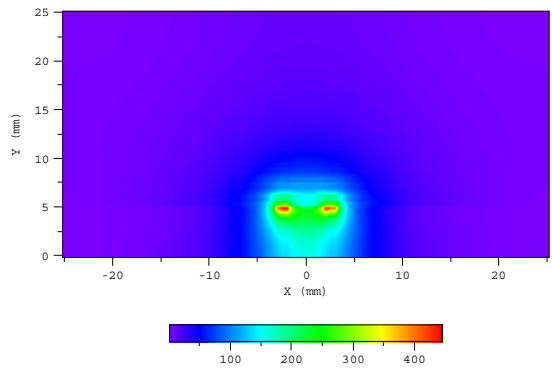

Fig. 17 (c)

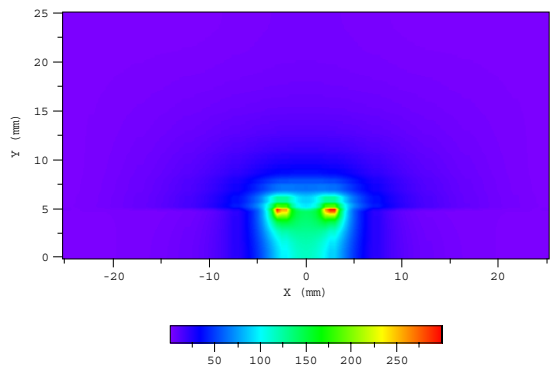

Fig. 17 (d)

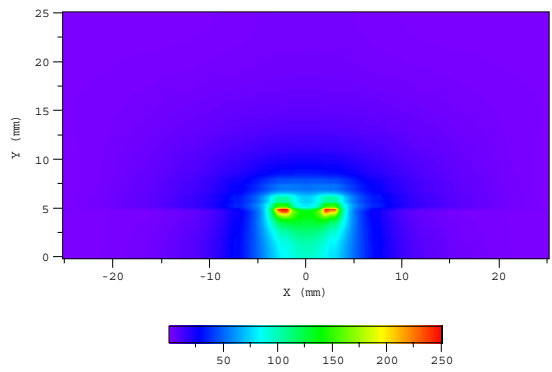

Fig. 17 (e)